\documentstyle[preprint,aps,tighten,eqsecnum]{revtex}

\begin{document}


\def \ts{\thinspace}
\def \beq{\begin{equation}}
\def \eeq{\end{equation}}
\def \beqa{\begin{eqnarray}}
\def \eeqa{\end{eqnarray}}
\def \EPS{\varepsilon}
\def \half{\hbox{$1\over2$}}
\def \zero{{\bf 0}}
 

\def \xf{x_{{}_F}}
\def \mn{m}
\def \GIII{G}
\def \Lbrk{\Bl\negthinspace\Bl}
\def \Rbrk{\Br\negthinspace\Br}
\def \LQCD{\Lambda_{\rm QCD}}
\def \Jt{\bbox{J}}
\def \qt{\bbox{q}}
\def \qz{q_{\scriptscriptstyle\parallel}}
\def \qIII{\vec{q}}
\def \xz{x_{\scriptscriptstyle\parallel}}
\def \xiz{\xi_{\scriptscriptstyle\parallel}}
\def \xit{\bbox\xi}
\def \xiIII{\vec{\xi}}
\def \xt{\bbox{x}}
\def \xIII{\vec{x}}
\def \xprimet{\xt^\prime}
\def \xprimeIII{\xIII\ts^\prime}
\def \Bl{\bbox{\Bigl[}}
\def \Br{\bbox{\Bigr]}}
\def \DIII{{\cal D}}
\def \SIII{{\cal S}}
\def \gnumX{ {{dN}\over{d\xf d^2\qt}} }
\def \gnumIII{ {{dN}\over{d\qz d^2\qt}} }
\def \Ihat{C}
\def \newscale{\kappa}


\def\.{\!\cdot\!}
\def\bk#1{\langle#1\rangle}
\def\bbk#1{\langle\!\langle#1\rangle\!\rangle}
\def\v#1{\bbox{#1}}
\def\u#1{^{(#1)}}
\def\l{\lambda}
\def\joint{Q}
\def\r{\tau}
\def\m{t}
\def\]{\right]}
\def\({\left(}
\def\[{\left[}
\def\){\right)}




\draft
\preprint{
  \parbox{2in}{McGill/01--02 \\
  hep-ph/0102337
}  }

\title{Non-Gaussian Correlations in the McLerran-Venugopalan Model}
\author{C.S. Lam$^*$ and 
Gregory Mahlon$^{*,\dagger}\phantom{\biggl\vert}$}
\address{$^*$Department of Physics, McGill University, \\
3600 University Street, Montr\'eal, Qu\'ebec  H3A 2T8 Canada \\[0.2cm]
$^{\dagger}$Department of Physics, University of Arizona, \\
1118 East Fourth Street, Tucson, AZ 85721 U.S.A.}
\date{February 27, 2001}
\maketitle
\begin{abstract}
We argue that the statistical weight function $W[\rho]$ appearing
in the McLerran-Venugopalan model of a large nucleus is intrinsically
non-Gaussian, even if we neglect quantum corrections.
Based on the picture where the nucleus of radius $R$ consists of 
a collection of color-neutral 
nucleons, each of radius $a \ll R$, 
we show that to leading order in $\alpha_s$ and $a/R$
only the Gaussian part of $W[\rho]$ enters into the final
expression for the gluon number density.  
Thus, the existing results in the literature which assume
a Gaussian weight remain valid. 
\end{abstract}


\section{Introduction}

With the first collisions at
the Brookhaven Relativistic Heavy Ion Collider,
the experimental frontier in both energy and density has 
advanced.  One of the goals of this machine is to search for
the so-called quark-gluon plasma (QGP), a deconfined state
of QCD where quarks and gluons form a sea of thermalized particles
occupying a volume of space significantly larger than that
of a typical hadron.
Theoretically, whether and when this occurs depends on the
initial conditions of the system.
That is, what is the distribution of the quarks and gluons
immediately before the nuclei collide?  

One particularly fruitful approach to the determination of parton
distribution functions in nuclei has been the 
McLerran-Venugopalan (MV) model\cite{paper1,paper2,paper3}.
The key observation underpinning the MV model is that for
large enough nuclei and for small enough values of the longitudinal
momentum fraction $\xf$, a new hard scale $\newscale^2$,
corresponding to the large color charge per unit transverse area,
enters into the problem.  
Consequently, when $\newscale^2 \gg \LQCD^2$, it is argued
that a classical treatment ought to provide a reasonable
approximation to the gluon distribution.
Three interrelated factors contribute to making the small-$\xf$ region
conducive to a classical description.  First, a large number of
color charges contribute to the source, implying that the vector
potential takes on large values within the nucleus.  
Second, 
the large vector potential corresponds to the existence of a large
number of gluons, inviting us to apply the Weisz\"acker-Williams
technique.  And, third, the new large scale appears in the
running strong coupling constant, $\alpha_s(\kappa^2)$, which
is small for sufficiently large $\kappa^2$.
In the classical treatment advocated in 
Refs.~\cite{paper1,paper2,paper3}, 
quantum mechanical expectation values are replaced by 
averages over a suitably chosen ensemble
of color sources whose statistical weight function, $W[\rho]$,
was argued to take on a Gaussian form.

Quantum corrections to the MV model were first computed
in Ref.~\cite{paper7}: they were found to be large
due to contributions of order $\alpha_s\ln(1/\xf)$.
The subsequent re-examination of the foundations of the MV model
performed in Ref.~\cite{paper9} led to the formulation
of a set of renormalization group equations (RGE) describing
the evolution of the weight function $W[\rho]$
as one changes the separation scale between hard and soft
partons\cite{paper11,paper39,paper40,paper43,%
paper34,paper44,paper91,paper96,paperXX}.
Early on in the development of the RGE
it was conjectured~\cite{paper9,paper11,paper39} that the
MV model could be viewed as an effective
theory at small $\xf$, with the effects
of the quantum corrections absorbed into
a renormalization of the weight function $W[\rho]$.
This conjecture has recently been proven by the
calculation in Ref.~\cite{paper91}.
While explicit solutions to the evolution equations for $W[\rho]$
have not yet been obtained, it is known that a
Gaussian distribution is {\it not}\ a solution to
these equations\cite{paper39}.

While these developments were taking place, we set out to address
a different issue, namely the poor infrared behavior
of the correlation functions in the MV model.
The two-point vector potential obtained in Ref.~\cite{paper9}
grows like $(x^2)^{x^2}$ at large distances ($x \agt \LQCD^{-1}$),
signaling the onset of non-perturbative effects associated with
confinement.  In Ref.~\cite{PAPER14} we observed that since
individual nucleons do not exhibit a net non-zero color charge, 
there should not be any long-range ($\gg \LQCD^{-1}$) correlations
between quarks.  This requirement of color neutrality
was cast into the form of
a mathematical constraint on the two-point charge density
correlation function.
We found that
the infrared divergence appearing in the MV model is completely
absent when the color neutrality condition is enforced\cite{PAPER14}.

With the infrared divergences under
control, it became clear that, at least in the classical
treatment of Ref.~\cite{PAPER14}, the $\xf$ dependence 
of the gluon distribution remained trivial, with 
$\xf dN/d\xf$ independent of $\xf$.
There are
two fairly obvious
possible sources of non-trivial $\xf$ dependence,
depending on the value of $\xf$ under consideration.
First, we can imagine taking $\xf$ to be somewhat larger than
the values specified in the MV model.  In this situation, 
the longitudinal resolution of the gluons becomes good enough
to start to probe the longitudinal structure of the Lorentz-contracted
nucleus which they see.  This avenue of investigation was
pursued in Ref.~\cite{PAPER16}, where we developed a fully
three-dimensional treatment of the classical gluon field
of a large nucleus.  
Our view of the nucleus (radius $R$) as containing 
color-neutral nucleons (each of radius $a$)
played a vital role in the calculation.
Taking advantage of the smallness of the
ratio $a/R$ for a large nucleus, and  
assuming that the nucleus possesses a
spherically symmetric distribution of color charge in its rest
frame, we found that the additional $\xf$ dependence 
manifests itself by the appearance of 
the combination $\qt^2 {+} (\xf \mn)^2$ in some (but not
all) of the functions parametrizing the final result
(see Eqs. (5.18)--(5.20) of Ref.~\cite{PAPER16}).
Here $\qt$ is the transverse momentum of the gluon and
$\mn$ is the  nucleon mass.  
Over most of the region where
our extended treatment is valid, $(\xf\mn)^2 \ll \qt^2$,
implying that these corrections are small.

The second possible source of non-trivial $\xf$ dependence,
namely quantum corrections
proportional to $\alpha_s \ln(1/\xf)$,
was proposed~\cite{paper1,paper2,paper3} even before 
the first calculation of these corrections in Ref.~\cite{paper7}
verified that they do indeed contain the necessary logarithms.
The idea is that a resummation of such powers 
to all orders would convert
the $1/\xf$ appearing in the leading-order
expression for $dN/d\xf$ into $1/\xf^{1+C\alpha_s}$
for some $C$.
One of the goals of the studies conducted in 
Refs.~\cite{paper9,paper11,paper39,paper40,paper43,%
paper34,paper44,paper91,paper96,paperXX} may thus be phrased
as the determination of $C$.
A priori it would appear that in order to determine the gluon
distribution in the interesting nonlinear regime, we would require
a complete solution to the evolution equations for $W[\rho]$
developed in Refs.~\cite{paper9,paper11,paper39,paper40,paper43,%
paper34,paper44,paper91,paper96,paperXX}.  And, furthermore,
since a Gaussian is not a solution to these equations, it would
seem that the general form of the gluon distribution obtained
in Refs.~\cite{paper9,PAPER14,PAPER16} could be fundamentally altered
once the non-Gaussian correlations are taken into account.
Such an outcome would be at odds with 
related calculations based on different 
techniques~\cite{paper21,paper19,paper20,%
paper63,paper64,paper66,paper67,paper27,paper37,%
paper12,paper15,paper30,paper65,paper92,paper48,%
paper58,paper41,paper59,%
paper72,paper73,paper69,paper70,paper71,paper85}
which already possess general agreement with the MV model.
The purpose of this paper is to investigate this issue.
Once again it will prove valuable to incorporate the physics
associated with confinement ({\it i.e.}\ the color neutrality
of the nucleons) into the discussion.  
We will demonstrate that to leading order in the strong coupling
and $a/R$ it is sufficient
to consider {\it only}\ the 
(renormalized) Gaussian contributions to the gluon
number density.   All contributions from the non-Gaussian 
portion of $W[\rho]$ are suppressed by additional factors
of $\alpha_s$ and/or $a/R$.  
Thus, the MV model remains in general agreement with the
various other approaches to small-$\xf$ physics referred to
above, even if we employ a non-Gaussian weight function.

In light of the arguments in favor of a Gaussian form for $W[\rho]$
put forth in Ref.~\cite{paper1}, the reader may be tempted to think
that our conclusion is a trivial consequence of the central
limit theorem.   However, as we will demonstrate, the central
limit theorem does not apply to this case.  Instead, the 
additional factors which suppress the non-Gaussian 
terms relative to their Gaussian counterparts come about
because of the fact that the individual nucleons
are color neutral objects which do not have significant 
color-charge correlations
at distances much greater than $a$.

The remainder of this paper is organized into two main sections.
In the first, we begin with a review of the central limit
theorem to establish the notation used in the subsequent discussion.
We will pay particular attention to 
the conditions under which the
central limit theorem is valid, and then show that for a large
nucleus these conditions are violated.  The essential observation
is that {\it any}\ non-trivial longitudinal correlations, no matter
how short range, spoil the Gaussian form of the weight function
in the very large nucleus ($A^{1/3}\rightarrow\infty$) limit.
In the second half of this paper we will return to the three-dimensional
framework for the MV model 
developed in Ref.~\cite{PAPER16} and consider the
additional terms which would be generated in the presence of
non-Gaussian contributions to $W[\rho]$.  Assuming only that the
nucleus consists of color-neutral nucleons of radius $a$ which
possess no nontrivial correlations at separations much larger than $a$,
we will demonstrate that no contributions beyond those already 
considered in Refs.~\cite{paper9,PAPER14,PAPER16} are generated
to leading order in $\alpha_s$ and $a/R$.  Thus, for a large enough 
nucleus, it
is sufficient to know only the two-point function associated with
$W[\rho]$, even when the quantum corrections are taken into account.
We close with a few words about the implications of our result.


\section{Gaussian and Non-Gaussian Distributions}
A Gaussian (or Normal) distribution in a random variable $\v{\m}$
usually comes about because of the {\it central limit theorem}. In this
section we review the conditions under which this theorem is valid,
and argue that all of these conditions are not,
in general, satisfied by
the color-charge distribution of a large nucleus, even when 
we omit radiative corrections. 
Hence, the statistical weight function $W[\rho]$
appearing in the MV model
is generically not expected to be a Gaussian, even classically.


\subsection{Generating Functions} 
Let $\v\r$ be a $D$-dimensional random variable, 
with normalized distribution $P(\v\r)$. For what
follows it is more convenient to specify it by its
moments:
\beq
\bk{\r_i\r_j\cdots\r_k}\equiv 
\int d^D\r\ \r_i\r_j\cdots\r_k \ts P(\v\r).
\label{mom-def}
\eeq
These two descriptions are equivalent because the Fourier transform
of $P(\v\r)$, namely
\beq
\widetilde P(\v\l)=
\int 
d^D\r \ts
e^{-i\v\l\,\.\,\v\r}
P(\v\r),
\eeq
has a Taylor series expansion in terms of the moments:
\beq
\widetilde P(\v\l)=1+
     \sum_{n=1}^\infty
{ {(-i)^n}\over{n!} }
\bk{(\v\r\cdot\v\l)^n}.
\label{ptilde}
\eeq

The moments defined by Eq.~(\ref{mom-def}) are reducible in
the following sense.
Suppose that the distribution $P(\v\r)$ factorizes as
\beq
P(\v\r) = \prod_{i=1}^D \ts p(\r_i),
\eeq
where $p(\r_i)$ is some single variable distribution,
{\it i.e.}\ suppose that the components of the vector $\v\r$
are completely unrelated to each other.  In this case,
the moment
$\bk{\r_i\r_j\cdots\r_k}$ factorizes into 
$\bk{\r_i}\bk{\r_j}\cdots\bk{\r_k}$.
More generally, however, the components of $\v\r$ will not be
completely independent of each other.  Then, the factorization
just described will fail: for example,
$\bk{\r_i\r_j} \ne \bk{\r_i}\bk{\r_j}$.
The difference between these two quantities provides
a measure of the amount of interdependence between the components
of $\v\r$.  
This information may be conveniently organized in terms
of the (irreducible) cluster moments $\bbk{\r_i\r_j\cdots \r_k}$,
which are recursively defined through the relations
\begin{mathletters}
\beqa
\bk{\r_i} &=& \bbk{\r_i},  \\ \cr
\bk{\r_i\r_j} &=& 
\bbk{\r_i\r_j}+\bbk{\r_i}\bbk{\r_j}, \\ \cr
\bk{\r_i\r_j\r_k}&=&
\bbk{\r_i\r_j\r_k}
+\bbk{\r_i}\bbk{\r_j\r_k}
+\bbk{\r_j}\bbk{\r_i\r_k}
+\bbk{\r_k}\bbk{\r_i\r_j}
+\bbk{\r_i}\bbk{\r_j}\bbk{\r_k}, \\ \cr 
\bk{\r_i\r_j\r_k\r_l} &=&
\bbk{\r_i\r_j\r_k\r_l}
\cr \phantom{\biggl[} &&
+\bbk{\r_i}\bbk{\r_j\r_k\r_l}+\bbk{\r_j}\bbk{\r_i\r_k\r_l} 
+\bbk{\r_k}\bbk{\r_i\r_j\r_l}+\bbk{\r_l}\bbk{\r_i\r_j\r_k}
\cr && 
+\bbk{\r_i\r_j}\bbk{\r_k\r_l}+\bbk{\r_i\r_k}\bbk{\r_j\r_l}
+\bbk{\r_i\r_l}\bbk{\r_j\r_k}
\cr \phantom{\biggl[} &&
+\bbk{\r_i}\bbk{\r_j}\bbk{\r_k\r_l}+\bbk{\r_i}\bbk{\r_k}\bbk{\r_j\r_l}
+\bbk{\r_i}\bbk{\r_l}\bbk{\r_j\r_k} 
\cr && 
+\bbk{\r_j}\bbk{\r_k}\bbk{\r_i\r_l}+
\bbk{\r_j}\bbk{\r_l}\bbk{\r_i\r_k}+
\bbk{\r_k}\bbk{\r_l}\bbk{\r_i\r_j}
\cr \phantom{\biggl[} &&
+\bbk{\r_i}\bbk{\r_j}\bbk{\r_k}\bbk{\r_l} . 
\eeqa
\end{mathletters}
The generalization to higher orders is obvious.
It turns out that the cluster moments are generated by 
$\ln\widetilde P(\v\l)$, 
in the same way that the ordinary moments are generated by 
$\widetilde P(\v\l)$:
\beq
\ln \widetilde P(\v\l) =
\sum_{n=1}^\infty
{ {(-i)^n}\over{n!} }
\bbk{(\v\r\cdot\v\l)^n}.
\label{lnptilde}
\eeq

A simple example is provided by the following Gaussian distribution:
\beq
P(\v\r) = { {\exp(-\v\r^2)}\over{{\pi}^{D/2}} }.
\label{special}
\eeq
All of the odd moments (reducible and irreducible) of Eq.~(\ref{special})
vanish.  Of the irreducible moments, only the two-point function
$\bbk{\r_i\r_j}$ is nonvanishing.  All other cluster moments are
zero, and the even reducible moments are expressible entirely
in terms $\bbk{\r_i\r_j}$.


\subsection{Central Limit Theorem}

Suppose the random variable $\v\r$ is {\it independently}
sampled $N$ times, yielding
the values $\v\r\u p\ (p=1,2,\ldots,N)$. The central limit theorem
asserts that when $N\gg 1$,
the mean value of these measurements,
\beq
\v{\m}
\equiv {1\over N} \sum_{p=1}^N\v\r\u p,
\eeq 
obeys a Gaussian
distribution with
\begin{mathletters} 
\beq
\bbk{\m_i}= {\bbk{\r_i}} = \bk{\r_i}
\eeq 
and
\beq
\bbk{\m_i\m_j}=
{ {\bbk{\r_i\r_j}} \over {N} }.
\eeq
\end{mathletters} 
In other words, the 
distribution of $\v{\m}$ is given by
\beq
W(\v{\m})=
\sqrt{{\det\Omega}\over{(2\pi)^D}}\ts
\exp\left[-\half(\m_i-\bk{\r_i})
           \Omega_{ij}(\m_j-\bk{\r_j})\right],
\label{gaussian}
\eeq
where
\beq
\(\Omega^{-1}\)_{ij}\equiv 
{ {\bbk{\r_i\r_j}} \over {N} }. 
\eeq

The proof of the central limit theorem
relies crucially
on the independence of samplings: 
the joint distribution $\joint$
of the sampled variables $\v\r\u p$ must be factorizable,
{\it i.e.}
\beq
\joint(\v\r\u 1,\v\r\u 2,\ldots,\v\r\u N)=\prod_{p=1}^NP(\v\r\u p).
\label{indep}
\eeq
In terms of $\joint$,
the probability $W(\v{\m})$ for finding the mean value
$\v{\m}$ is given by
\beq
W(\v{\m})=
\int d^D\v\r\u 1 
\int d^D\v\r\u 2 
\cdots 
\int d^D\v\r\u N \ts 
\joint(\v\r\u 1,\v\r\u 2,\ldots,\v\r\u N)\ 
\delta^D \biggl(\v{\m}-
               \sum_{p=1}^N\v\r\u p/N\biggr).
\label{w}
\eeq
Using the 
distribution given
in~(\ref{indep}), the moment generating function is simply
\beqa
\widetilde W(\v\l)&=&
\int d^D\v{\m} \ts\ts
e^{-i\v\l\cdot\v{\m}}\ts W(\v{\m})
\cr &=&
\[\widetilde P\({\v\l / N}\)\]^N.
\eeqa
This leads to the cluster moment generating function
\beqa
\ln \widetilde W(\v \l) &=&
N \ln\widetilde P\({\v\l / N}\)
\nonumber\\[0.05cm] &=&
\sum_{n=1}^\infty 
{ {(-i)^n} \over {n!} }
{
 { \bbk{(\v\r\.\v\l)^n} }
\over
 { \ts N^{n-1} }
},
\label{lnw}
\eeqa
where we have applied Eq.~(\ref{lnptilde}) to obtain the
second line.
Now consider the $N\gg 1$ limit.
If we approximate~(\ref{lnw}) by just the first term,
we obtain $W(\v\l)=\exp(-i\v\l\cdot\langle\v\r\rangle)$,
corresponding to $W(\v\m)=\delta^D(\v\m-\bk{\v\r})$.
If instead, we include the first two terms in the infinite series,
we get
\beq
\widetilde W(\v\l)
=\exp\Bigl(-i\v\l\.\bk{\v\r}-\l_i\l_j\bbk{\r_i\r_j}/2N\Bigr),
\eeq
corresponding to the Gaussian distribution~(\ref{gaussian}).
Deviations from this Gaussian are suppressed by the additional
powers of $1/N$ present in the higher order terms.


\subsection{Non-Gaussian Distribution}\label{NONGAUSS}
The proof of the central limit theorem 
depends critically on the assumption of independent sampling.
If~(\ref{indep}) is violated, then the proof fails.
Even a small correlation between successive samplings
is sufficient to destroy the conclusion, as the following example 
illustrates.

Suppose we introduce a ``nearest-neighbor'' correlation by 
replacing~(\ref{indep}) with
\beq
\joint(\v\r\u 1,\v\r\u 2,\ldots,\v\r\u N)
=
\Biggl[
1 +
\sum_{m=1}^{N-1}
{
{ f(\v\r\u m,\v\r\u {m+1}) }
\over
{ P(\v\r\u m) \ts P(\v\r\u{m+1}) }
}
\Biggr] 
\prod_{p=1}^NP(\v\r\u p).
\label{correl}
\eeq
To ensure
that~(\ref{correl}) is properly normalized,
we further assume  $f(\v\r\u m,\v\r\u{m+1})$ 
to be symmetric in its arguments, and that
\beq
\int d^D\v\r\u mf(\v\r\u m,\v\r\u{m+1})=0.
\eeq
The moment generating function
for the distribution~(\ref{w}) is now
\beq
\widetilde W(\v\l)=
\Bigl[
{\widetilde{P}} (\v\l / N)
\Bigr]^N
\left\{1+{
{
{ (N-1) \widetilde f(\v\l) }
\over
{ \Bigl[{\widetilde{P}}(\v\l / N)\Bigr]^2} }
}
\right\},
\label{newwt}
\eeq
where
\beq
\widetilde f(\v\l)\equiv
\int d^D\v\r\u 1
\int d^D\v\r\u 2 \ts
e^{-i\v\l\cdot[\v\r\u 1+ \v\r\u 2]} \ts
f(\v\r\u 1,\v\r\u 2).
\eeq
Note that $\widetilde f(\v 0)=0$.
Unlike Eq.~(\ref{lnw}),
the irreducible moments generated by
\beq
\ln\widetilde W(\v\l) =
N\ln\widetilde P(\v\l / N)
+\ln\left\{1+{
{
{ (N-1) \widetilde f(\v\l /N) }
\over
{ \Bigl[{\widetilde{P}}(\v\l / N)\Bigr]^2} }
}
\right\}
\label{newc}
\eeq
do not approach a linear function of $\v\l$ as $N\to\infty$.
In fact, we have
\beq
\ln\widetilde W(\v\l)=-i\v\l\.\bk{\v\r}+\ln\[1-i\v\l\.\v\phi\]
+{\cal O}(1/N),
\label{nlimit}
\eeq
where
\beq
\phi_j \equiv i{{\partial}\over{\partial\lambda_j}}\ts 
\widetilde f(\v\l)\biggl\vert_{\lambda=0}.
\eeq
For $\v\phi\not=0$, Eq.~(\ref{nlimit}) is nonlinear in $\v\l$.
This distribution possesses non-zero values for {\it all}
of its irreducible moments, independent of $N$.
Including the ${\cal O}(1/N)$ contributions does not
change this conclusion.  The corresponding 
distribution $W(\v\m)$ is no longer a pure Gaussian:
the central limit theorem does not hold in this case.
While it is straightforward to determine the (new) form of
the distribution function 
for this particular example, in general this task will not be
so easy.  Fortunately, all we need to know for what follows is the
fact that the resulting
distribution is fundamentally non-Gaussian when the successive
measurements are (even mildly) correlated.

It is useful to see a bit more about
how the correlators change when we relax the assumption
of independent measurements.
With independent sampling, as in Eq.~(\ref{indep}),
correlators such
as $\bk{\r_i\u p\r_j\u q\cdots \r_k\u s}$
factorize into
groups with the same superscripts. For example, 
\beqa
\bk{\r_i\u p\r_j\u q \r_k\u s\r_l\u s} &=&
\int d^D\v\r^{(1)} \cdots
\int d^D\v\r^{(N)} \ts\ts
\r_i^{(p)} \r_j^{(q)} \r_k^{(s)} \r_l^{(s)} \ts
P(\v\r^{(1)}) \cdots P(\v\r^{(N)}) 
\cr \phantom{\biggl[} 
&=& \bk{\r_i}\bk{\r_j}\bk{\r_k\r_l},
\label{fact}
\eeqa
assuming that $p$, $q$, and $s$ all take on distinct values.
Of particular interest 
in connection with the MV model will be the special case
\beq
\langle \r_i^{(p)} \r_j^{(q)} \rangle =
\delta^{pq} \langle \r_i \r_j \rangle
+(1-\delta^{pq}) \langle \r_i \rangle \langle \r_j \rangle.
\label{relevant}
\eeq
In contrast, if the measurements are not independent, such
as in~(\ref{correl}),
then the factorizations obtained in Eqs.~(\ref{fact})
and~(\ref{relevant}) are no longer valid.
Additional non-factorizable correlators involving different
superscripts will, in general, be present,
signaling the breakdown of the central limit theorem.


\subsection{Color Charge Distribution in a Large Nucleus}\label{CONNECT}

Let $\rho^a(\xz,\xt)$ be the color charge density of
a large nucleus with color $a$ and atomic weight $A\gg 1$, at
the transverse position $\xt$
and the longitudinal position $\xz$ (a precise definition
of $\xz$ is given in Eq.~(\ref{xl-def}) below). 
Let 
\beq
\rho^a_2(\xt)\equiv \int_{-\infty}^{\infty} d\xz\ts \rho^a(\xz,\xt)
\eeq 
be the accumulated
two-dimensional charge density. 
On account of Lorentz contraction
of the longitudinal size of the relativistic nucleus,
this is expected to be the relevant quantity
governing the gluon distribution, provided that
only valence quarks are included in $\rho$, and provided that
the longitudinal wavelength of the gluon is much bigger than the 
longitudinal size of the relativistic nucleus. 
These are two of the basic assumptions underlying the MV
model\cite{paper1,paper2,paper3}.

The typical size of
$\rho^a_2(\xt)$ for a nucleus is of order $A^{1/3}$ times 
the typical size of $\rho^a_2(\xt)$ for a 
single nucleon. 
If we identify $A^{1/3}$ with the number of measurements $N$ of the 
previous subsections, 
and the accumulated charge density $\rho^a_2(\xt)/A^{1/3}$ with the 
mean value $\m_i$ (with $i\to \{a,\xt\}$), 
then the central limit
theorem would force $W(\v{\m})$ 
(that is, $W[\rho_2(\xt)]$) to be Gaussian 
in the limit of large $A^{1/3}$, provided that
the conditions needed to prove
the theorem are obeyed. 
With this mapping, the superscript $p$ in the 
random variable $\r_i\u p$ corresponds to $\xz$,
and $\r_i\u p$ corresponds to $\rho^a(\xz,\xt)$.
A summary of these connections appears in Table~\ref{Translation}.

As discussed in Sec.~\ref{NONGAUSS}, a necessary condition for the
central limit theorem to hold is that the correlators
$\bk{\r_i\u p\r_j\u q\cdots \r_k\u s}$ must factorize into groups
with identical superscripts. 
In particular,  consider the translation of Eq.~(\ref{relevant})
into the MV model.  The color neutrality condition tells us that
$\bk{\rho^a(\xz,\xt)}=0$.  Therefore, we are left with  
\beqa
\bbk{\rho^a(\xz,\xt)\rho^b(\xz',\xprimet)} &=&
\bk{\rho^a(\xz,\xt)\rho^b(\xz',\xprimet)} \cr
&=& \delta(\xz-\xz'\ts)
\langle \rho_2^a(\xt) \rho_2^b(\xprimet) \rangle.
\label{factd}
\eeqa
This is effectively the form of correlator argued for in
Ref.~\cite{paper1}, and employed extensively in the
literature~\cite{paper2,paper3,paper7,paper9,paper91,PAPER14,%
paper12,paper4,paper5,paper6,paper24,paper10,paper26,paper79}.

However, there are two reasons why it is desirable to go
beyond the Gaussian approximation represented by Eq.~(\ref{factd}).
First, at very small $\xf$ the quantum corrections in the MV model
become large.   If we wish to incorporate these corrections
via the RGE analysis of 
Refs.~\cite{paper9,paper11,paper39,paper40,paper43,%
paper34,paper44,paper91,paper96,paperXX}, we must go beyond
the Gaussian approximation, since a Gaussian is {\it not}\ a solution
to these equations\cite{paper39}.
Second, if we go to somewhat larger values of $\xf$, where
the quantum corrections may not be too large,
the presence of longitudinal correlations between
quark charges within a
nucleon and the fact that the
longitudinal size of a nucleon is not exactly
zero begin to have an effect, even classically.
In Ref.~\cite{PAPER16}, we employ the form
\beq
\bbk{ \rho^a(\xz,\xt) \rho^b(\xz',\xprimet) }
\propto \delta^{ab}
\Bigl[\delta(\xz-\xz')\ts\delta^2(\xt-\xprimet) 
- \Ihat(\xz-\xz';\xt-\xprimet)\Bigr].
\label{rhorho3}
\eeq
where $\Ihat(\xz-\xz';\xt-\xprimet)$ is a
reasonably smooth function parameterizing the
mutual correlations between pairs of quarks.   
This function enforces the color neutrality condition.
The nonfactorizability of Eq.~(\ref{rhorho3}) tells us
that for a nucleus with a non-zero longitudinal
thickness we should not
expect any single layer to be color neutral by itself.
Equivalently, we may view the color neutrality condition
as forcing us to have non-trivial longitudinal correlations.
In Ref.~\cite{PAPER16},
all of the higher-order cluster moments of the
distribution corresponding to Eq.~(\ref{rhorho3}) are
taken to vanish,
implying that $W[\rho^a(\xz;\xt)]$ is Gaussian.
However,
Eq.~(\ref{factd}) is violated, and
the corresponding 2-dimensional distribution
$W[\rho^a_2(\xt)]$ is {\it not}\ a Gaussian, because
the central limit theorem does not apply to this computation.
Nevertheless, the result obtained in Ref.~\cite{PAPER16} 
remarkably 
turns out to have exactly the same form as if $W[\rho^a_2(\xt)]$
had been Gaussian:
for sufficiently
small $\xf$, it matches the results of Refs.~\cite{paper9,PAPER14}.  
In the next section we will 
see that this outcome is quite general, by 
considering the consequences of allowing for a
non-Gaussian weight function.


\section{Effect of Non-Gaussian Contributions}

We have just demonstrated that
we do not expect the statistical weight
function $W[\rho_2(\xt)]$ to have a Gaussian form, even
in the purely classical case.  On the other hand, a Gaussian
has been frequently employed in the literature dealing with the MV 
model~\cite{paper1,paper2,paper3,paper7,paper9,paper91,PAPER14,%
paper12,paper4,paper5,paper6,paper24,paper10,paper26,paper79}.
In this section we will argue that for a large enough nucleus,
the results which would be obtained with a non-Gaussian weight
function are identical to those which follow from a Gaussian 
weight function to leading order in $\alpha_s$ and $A^{1/3}$.
That is, the new contributions introduced by the non-Gaussian
weight are suppressed by additional factors of $\alpha_s$ and/or
do not contain as many factors of $A^{1/3}$.
Our conclusion is a consequence of the  color neutrality of
the nucleons:  all non-trivial correlations are limited
to length scales of order $a \sim \LQCD^{-1}$.
Beyond the color neutrality of individual nucleons, we will
assume nothing about the form of $W[\rho(\xIII\ts)]$.
Our discussion is framed in terms of the three-dimensional
extension of the MV model introduced in Ref.~\cite{PAPER16},
with the addition of non-zero higher-point cluster moments.


\subsection{Diagrammatic Representation of the Gluon Number Density}

Imagine a very large nucleus as viewed in its rest frame.
The current corresponding to this situation may be written
in the simple form
\beq
J_r^0 = \rho(-z_r,\xt_r); \qquad
J_r^1 = J_r^2 = J_r^3 = 0,
\label{Jrest}
\eeq
where we  employ the subscript ``$r$'' to denote rest frame
quantities.  In this frame the Yang-Mills equations for 
the vector potential possess the
``obvious'' time-independent Coulomb solution.
Furthermore, since only $A_r^0 \ne 0$, we have
$\partial_0 A^{0}_r = \partial \cdot A_r =0$:
the Coulomb solution is  the same as the covariant
gauge solution in this frame.  
Thus, we conclude that
even when we boost to the lab frame, it is natural to begin with the
covariant gauge solution for the vector potential\cite{PAPER16}.  
Indeed, it has been observed that although the expression for
the gluon number density is most easily written in terms of the
vector potential in the light-cone gauge
(see Eq.~(\ref{gnum3d}) below), it is nevertheless
easiest to work in terms of the covariant gauge expression for the 
current\cite{paper91,paper96,paper18,paper12}.

In the lab frame, where the nucleus is moving along the $+z$ axis
with speed $\beta$, we take the source to
be of the form~\cite{PAPER16}
\beq
J^{+} = {{1}\over{\EPS}} \ts 
\rho( \xz, \xt );
\qquad J^{-} = {{\EPS}\over{2}} J^{+};
\qquad \Jt = \zero.
\label{Jlab}
\eeq
The longitudinal coordinate $\xz$ is defined by
\beq
\xz \equiv 
{{1}\over{\EPS}}\ts x^{-} 
- {{\EPS}\over{2}}\ts  x^{+}.
\label{xl-def}
\eeq
The parameter
\beq
\EPS \equiv \sqrt{ { {2(1-\beta)} \over {1+\beta} }}
\label{epsdef}
\eeq
measures how close the source is to being
exactly on the light 
cone.\footnote{We define the light-cone coordinates to be 
$x^{\pm} = -x_{\mp} = (x^0 \pm x^3)/\sqrt{2}.$
The transverse coordinates $x^1$ and $x^2$ form a two-vector 
which we write in bold-face: $\xt$.  
Our metric has the 
signature $({-},{+},{+},{+})$.  Thus, the scalar product in 
light-cone coordinates reads
$ q_{\mu} x^{\mu} = -q^{+}x^{-} - q^{-}x^{+} + \qt\cdot\xt. $}
As explained in Ref.~\cite{PAPER16}, we prefer to work with
natural (order unity) quantities, and keep track of all small
and large parameters explicitly through $\EPS$.
At the end of the calculation we let $\EPS\rightarrow 0$.
Interestingly, in terms of $\xz$, the function describing
the source is still spherical:  the Lorentz contraction that
shrinks $x^{-}$ towards zero in the lab frame is exactly compensated
by the factor $1/\EPS$.   
Reflecting this fact, we will frequently use the notation
$\xIII \equiv (\xz;\xt)$.

As explained above, it is natural to first solve 
the Yang-Mills equations for the vector
potential in the covariant gauge, using the covariant gauge
source~(\ref{Jlab}), and to subsequently transform the result
into the light-cone gauge $A^{+} = 0$.
The result of this procedure may be written in the form
\beqa
A^{j}(\qz;\xt) = 
{{g}\over{i\qz}} \sum_{m=1}^{\infty}
(-ig^2)^{m-1}  &&
\int_{-\infty}^{\infty}  d^m{\xz}_{\downarrow} \ts \exp(-i\qz{\xz}_1)
\int d^3\xiIII_1 \ts
\partial^j\GIII({\xz}_1-{\xiz}_1;\xt-\xit_1)
\cr  \times &&
\Biggl(\ts\prod_{k=2}^{m}\int d^3\xiIII_k \ts
\GIII({\xz}_k-{\xiz}_k;\xt-\xit_k)\Biggr)
\Lbrk
\rho(\xiIII_1)
\rho(\xiIII_2) \cdots 
\rho(\xiIII_m)
\Rbrk,
\label{LCA}
\eeqa
where $\qz$ is the momentum conjugate to $\xz$, {\it i.e.}
\beq
A^j(\qz;\xt) \equiv 
\int_{-\infty}^{\infty} d\xz \ts
\exp(-i\qz\xz) \ts A^j(\xz,\xt).
\label{FTA}
\eeq
The power counting rules we will present in Sec.~\ref{POWERS}
are simplified by using $A(\qz;\xt)$ instead of $A(\xz,\xt)$.

\begin{figure}[h]

\vspace*{17cm}
\includegraphics{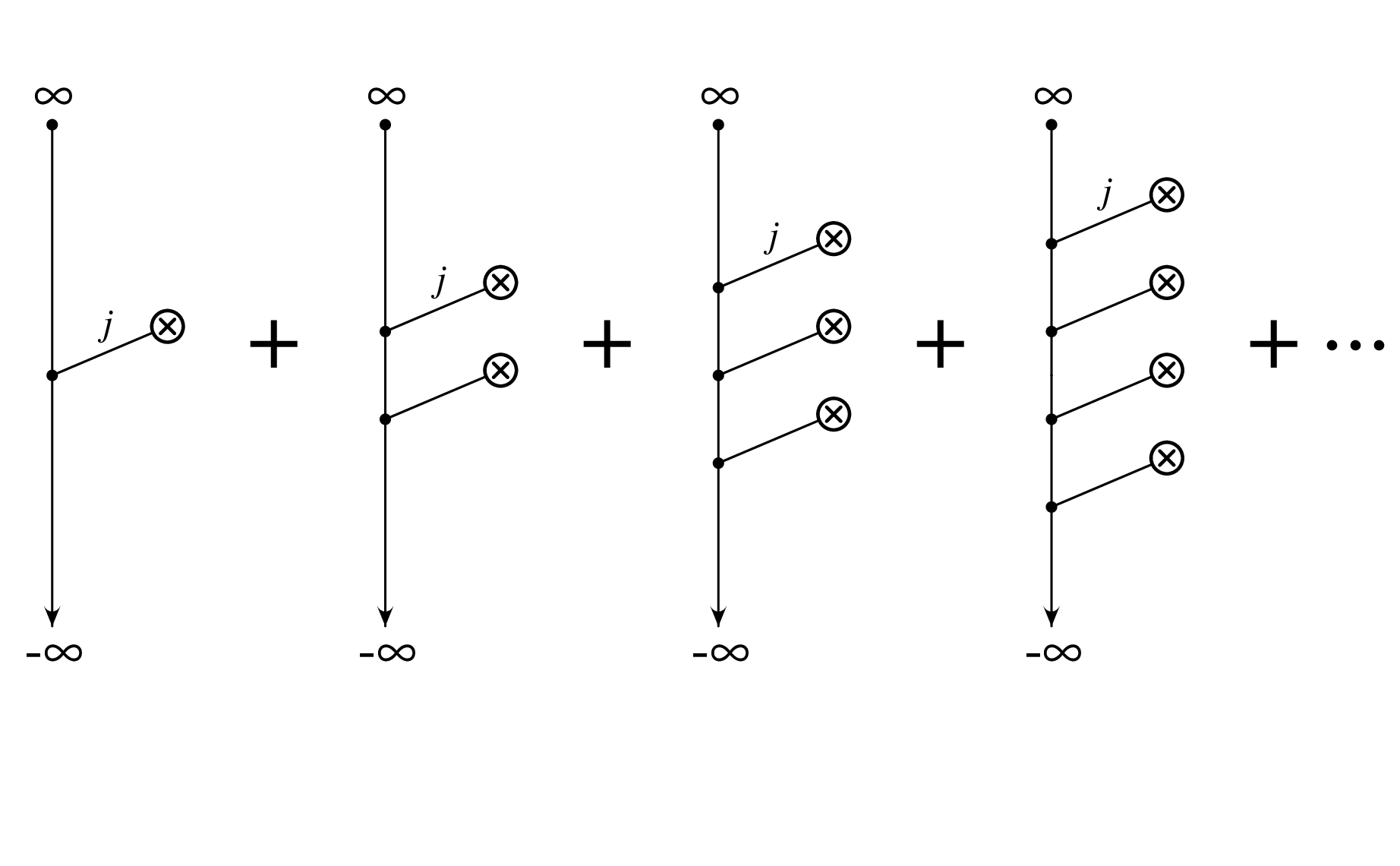}

\caption[]{Diagrammatic representation of the series
expansion for the light-cone gauge vector potential,
Eq.~(\protect\ref{LCA}).  The circled crosses denote the
positions at which the sources are being evaluated.  The dots
represent the ordered integrations coming from the gauge transformation
into the light-cone gauge.  A propagator connects each of these points
to the sources.  
}
\label{DiagA1}
\end{figure}

Although Eq.~(\ref{LCA}) appears complicated
at first glance, it is easily understood in terms of the
diagrammatic representation introduced in Ref.~\cite{PAPER16}.
The light-cone gauge vector potential is a non-linear function
of the source, containing all possible ``powers'' of $\rho$.
Note that the series begins at order $g$, and that each additional
occurrence of the source adds a factor of $g^2$.
Fig.~\ref{DiagA1} illustrates the first few diagrams corresponding
to the series.
The $m$th order diagram consists of $m$ copies of the source, 
represented by the circled crosses.  In Eq.~(\ref{LCA}) they appear
as the nested multiple commutator
\beq
\Lbrk
\rho(\xiIII_1)
\rho(\xiIII_2)  \cdots 
\rho(\xiIII_m)
\Rbrk
\equiv 
\Bl\Bl\Bl \cdots \Bl  \rho(\xiIII_1),
                      \rho(\xiIII_2)\Br,
                      \rho(\xiIII_3)\Br,
          \cdots \Br, \rho(\xiIII_m)\Br.
\eeq
To each source we attach a propagator (Green's function) 
connecting the source point
$({\xiz}_k,\xit_k)$ to the point $({\xz}_k,\xt)$:
\beq
\GIII({\xz}_k{-}{\xiz}_k; \xt{-}\xit_k) =
-{ {1}\over{4\pi} } \ts
{ {1}\over\sqrt{({\xz}_k-{\xiz}_k)^2 + (\xt-\xit_k)^2} }.
\label{propagator}
\eeq
Here
$\xt$ is the transverse position at which  we wish to know the
vector potential.  The index labelling the uppermost propagator
on the diagram represents
the derivative indicated in Eq.~(\ref{LCA}).
Finally, we have an ordered integration over the longitudinal
variables ${\xz}_k$:
\beq
\int_{-\infty}^{\infty} d^m{\xz}_{\downarrow}
\equiv
\int_{-\infty}^{\infty} d{\xz}_1
\int_{-\infty}^{{\xz}_1} d{\xz}_2
\cdots
\int_{-\infty}^{{\xz}_{m-1}} d{\xz}_m.
\eeq
Mathematically, this integration results from the gauge
transformation to the light-cone gauge from the covariant gauge.
Physically this integration corresponds to the final state
rescatterings which would be present in a computation
of the gluon number density based entirely on the covariant
gauge\cite{paper15,paper30,paper65,paper92}.
The ordered integration is represented by the dots on the
vertical line.  The dots are to slide up and down the entire
length of the line without passing one another.
Although we have written all of the integrations over an
infinite range, in practice the source provides non-zero contributions
only over a region of size $R$, the nuclear radius.


The next step is to connect the gluon number density to the
two-point correlation function for the vector 
potential\cite{PAPER16,CollinsSoper}:
\beqa
\gnumIII &\equiv &
{ {\qz}\over{4\pi^3} } \ts
\int d^2\xt
\int d^2\xprimet \ts
e^{i\qt\cdot(\xt-\xprimet)}
\langle A_i^a(\qz;\xt) A_i^a(-\qz;\xprimet) \rangle.
\label{gnum3d}
\eeqa
Intuitively, the result in Eq.~(\ref{gnum3d})  may be understood
by envisioning the expansion of the vector potential in terms
of creation and annihilation operators and recognizing
that $\langle AA\rangle$ contains the number 
operator\cite{paper3,paper4}.
The vector potential appearing in Eq.~(\ref{gnum3d}) is 
in the light-cone gauge.
This choice reflects the fact
that the intuitive picture of the parton model is 
most transparently realized in the light-cone 
gauge\cite{CollinsSoper,Curci,LC1,paper31}.
Strictly speaking, the angled brackets in Eq.~(\ref{gnum3d}) 
represent a quantum-mechanical expectation value.  In the MV
model, we make a classical approximation to
this quantity by performing an ensemble average with
an appropriate weighting function $W[\rho(\xIII\ts)]$.  

Given a specific form for $W[\rho(\xIII\ts)]$, we may 
(in principle) evaluate Eq.~(\ref{gnum3d}) by inserting two copies
of~(\ref{LCA}), performing the required average term-by-term,
and summing the resulting series.
Diagrammatically, the quantity $\langle AA \rangle$
may be represented by drawing all
possible pairs of diagrams for a single $A$,
with the understanding that all possible contractions should be
performed
(see Figs.~\ref{SecondOrder}--\ref{FourthOrder} for the
first three orders in this expansion).  For a Gaussian
weight function, only pairwise contractions appear.
On the other hand, for a non-Gaussian distribution, contractions
connecting three or more sources must also be considered.
Because the nucleons are neutral, 
$\langle \rho(\xIII\ts) \rangle = 0$.
Hence there must be no uncontracted sources left over in any
diagram.


\subsection{Power Counting Rules}\label{POWERS}

To proceed further, we must make some reasonable assumptions
about the correlation functions corresponding to the moments
of $W[\rho(\xIII\ts)]$.    
Let us parameterize the $m$-point correlation function
($m>1$) by
\beq
\bbk{\rho(\xIII_1)\rho(\xIII_2)\cdots\rho(\xIII_m)}
\equiv
\SIII_m\biggl( { {1}\over{m} } \sum_{j=1}^m \xIII_j \biggr) \ts
\DIII_m( \xIII_1{-}\xIII_2; 
         \xIII_2{-}\xIII_3; \ldots ; 
         \xIII_{m-1}{-}\xIII_m ).
\label{multi-rho}
\eeq
The exact choice made for the $m{-}1$ difference coordinates
is arbitrary and unimportant to our argument.  What does
matter is the fact that the source 
corresponds to a large nucleus of radius $R$ which is in turn
composed of $A$ nucleons, each of radius $a$.  The nucleons
themselves are color-neutral.  Because of confinement, we
do not expect the field to be correlated at distances much greater
than $a$:  what is happening inside one nucleon is largely
independent of what is happening inside of the others.
Thus, the function $\DIII_m$ ought to be small unless 
$\vert\xIII_i - \xIII_j \vert \alt a$ for all pairs of points.
Furthermore, the center-of-mass coordinate ought to point
at a position somewhere inside the nucleus ({\it i.e.}\ it
should have a magnitude $\alt R$)
in order for $\SIII_m$ to take on a non-negligible value.

The above physical considerations are sufficient to allow
us to determine the order of magnitude of an arbitrary
diagram in terms of powers of $\alpha_s$ and 
$R/a$ ($A^{1/3}$).
The powers of the coupling are simple.  
Recall that each of the diagrams
for the gluon number density~(\ref{gnum3d}) are formed by
gluing together two copies of the expansion for the 
vector potential~(\ref{LCA}).  Since~(\ref{LCA}) contains only
odd powers of $g$, the diagram representing the contribution
to~(\ref{gnum3d})  with a total of $j$
sources contains the factor $g^{2j-2}$
(or, equivalently $\alpha_s^{j-1}$).

Now let us consider the integrations which go into the computation
of the contributions to 
$\langle A_i^a(\qz;\xt) A_i^a(-\qz;\xprimet) \rangle$.
The integrand for a given diagram will contain several propagators
plus factors of $\SIII_m$ and $\DIII_m$, depending upon how the
sources are contracted.  As described above, $\DIII_m$ contains
the length scale $a$ whereas $\SIII_m$ contains the length scale $R$.
The Green's functions contain no intrinsic scale of their own.
What we would like to know is how many powers of $R/a \sim A^{1/3}$
are generated when we perform the required integrations,
which range over all
possible locations of the sources as well as the entire length
of the two vertical lines.  

\begin{figure}[h]

\vspace*{9.5cm}
\includegraphics{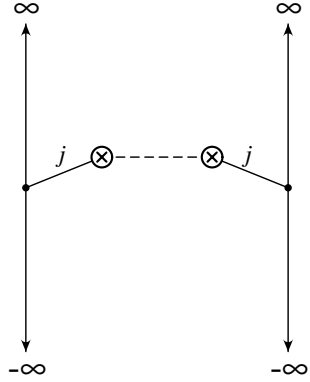}

\caption[]{The lowest-order contribution to the gluon number density,
Eq.~(\protect\ref{gnum3d}).  
According to the power-counting rules
described in the text, this diagram is proportional to $\alpha_s A$.
}
\label{SecondOrder}
\end{figure}

\begin{figure}[h]

\vspace*{9.5cm}
\includegraphics{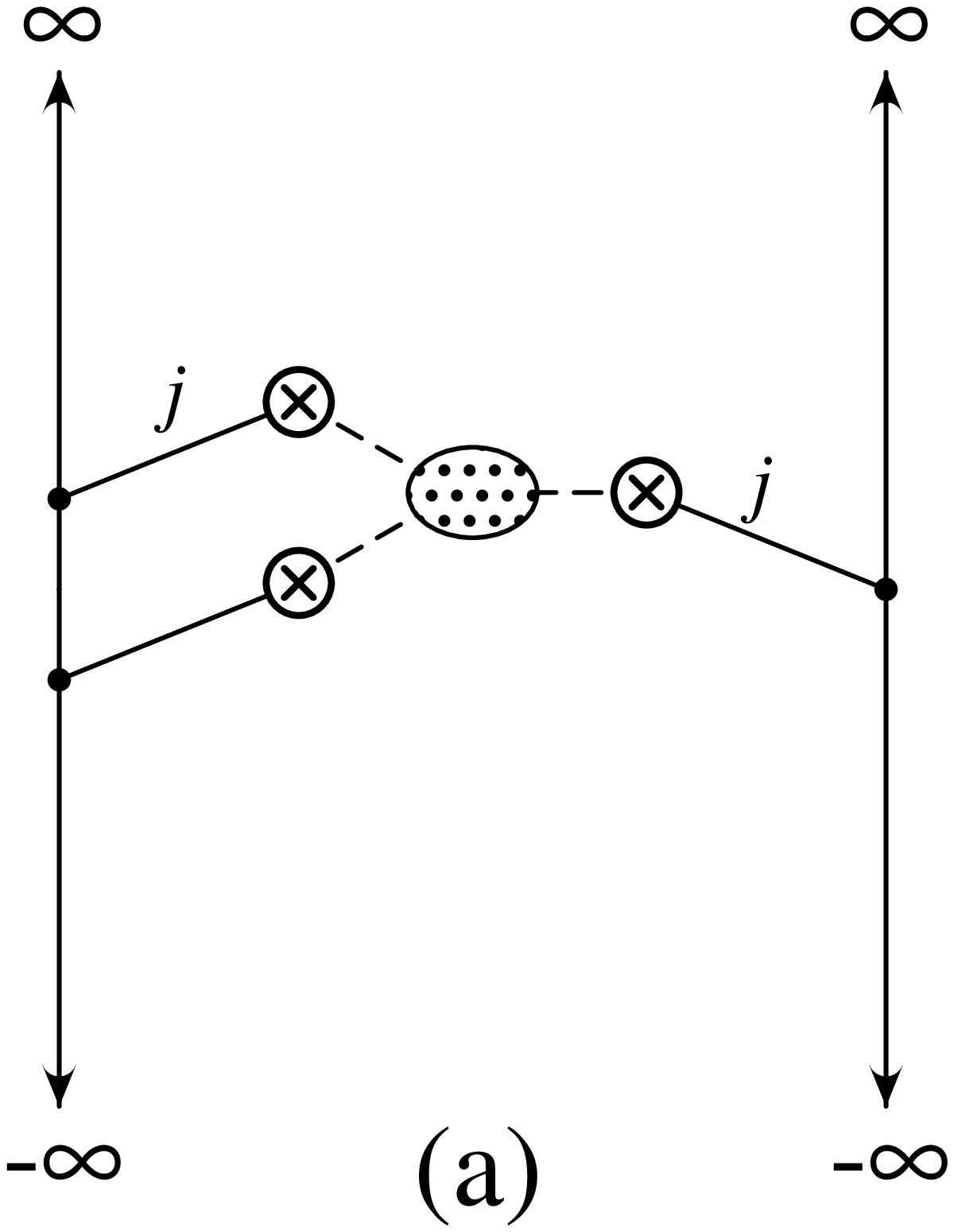}
\includegraphics{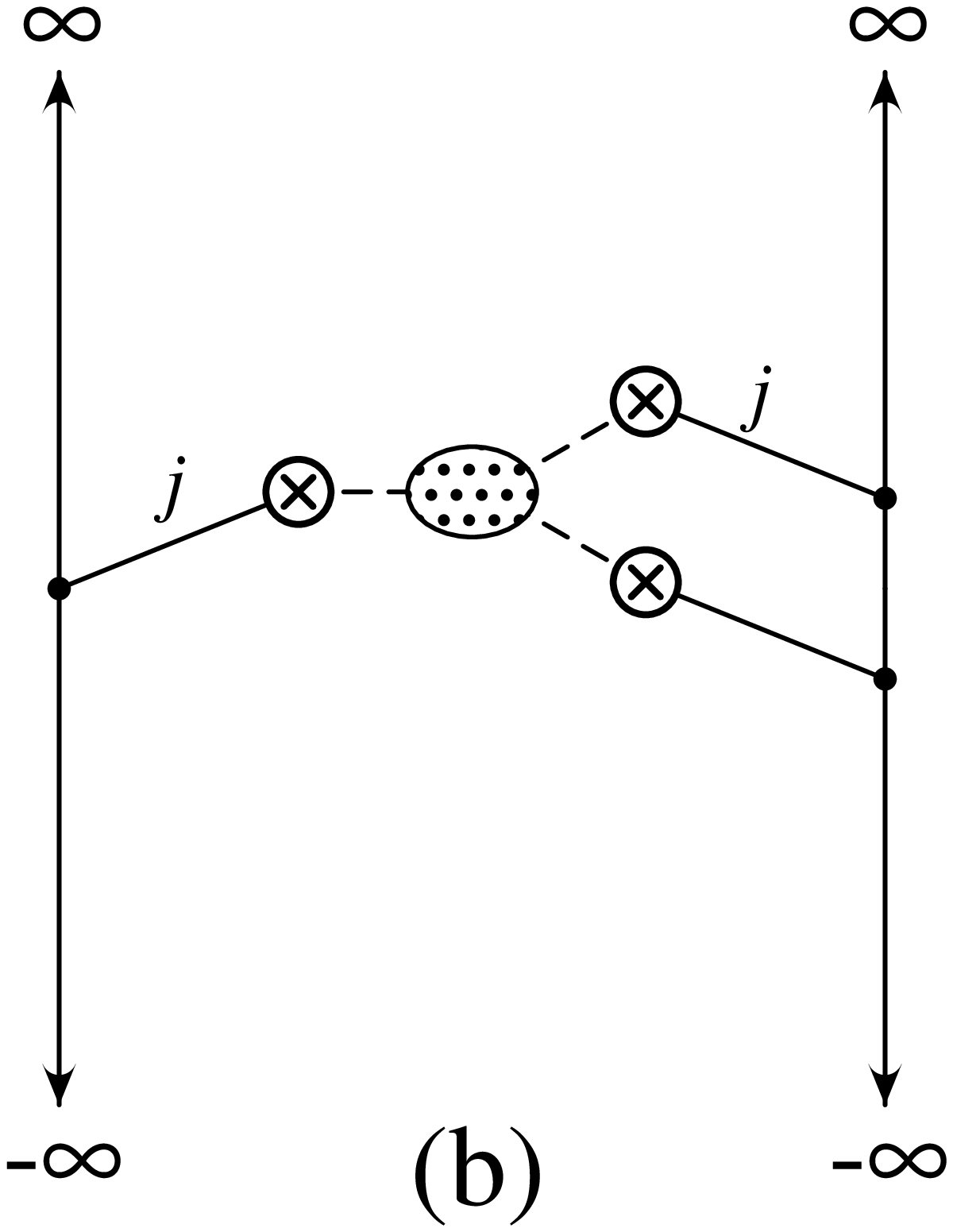}

\caption[]{The contributions to the gluon number 
density~(\protect\ref{gnum3d})
at 3rd order in $\rho$. 
According to the power-counting rules
described in the text, both diagrams are proportional to $\alpha_s^2 A$.
}
\label{ThirdOrder}
\end{figure}

\begin{figure}[h]

\vspace*{20cm}
\includegraphics{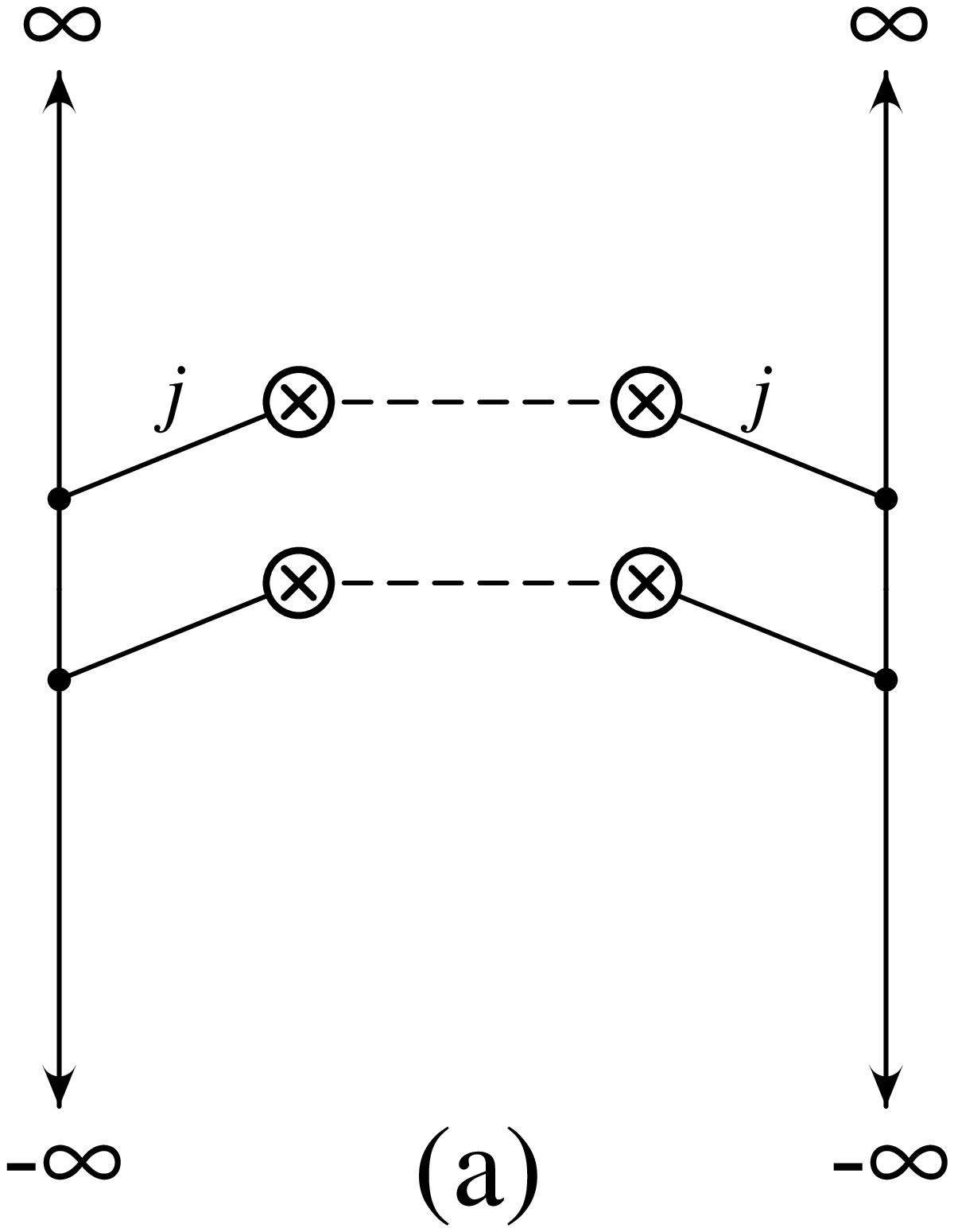}
\includegraphics{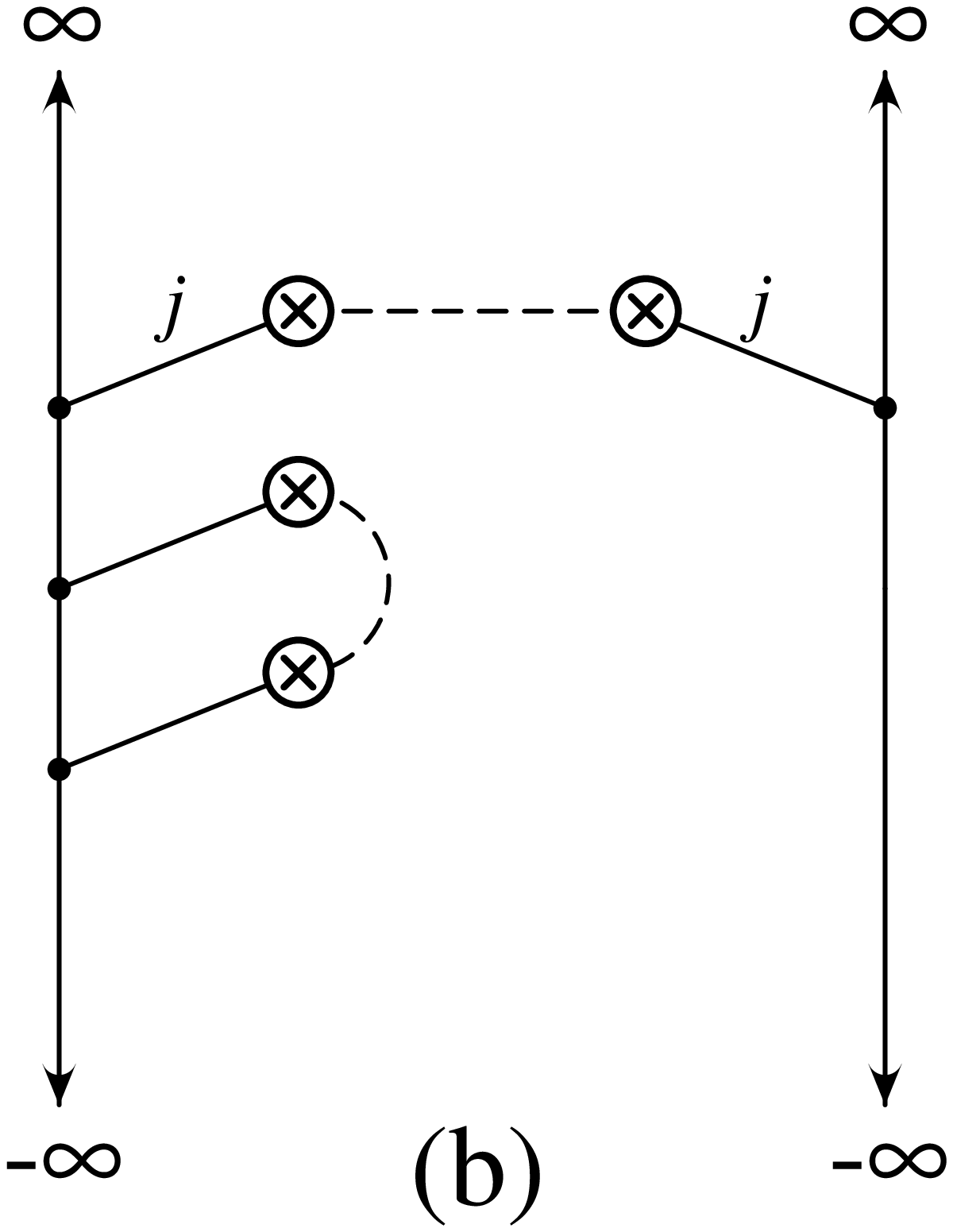}
\includegraphics{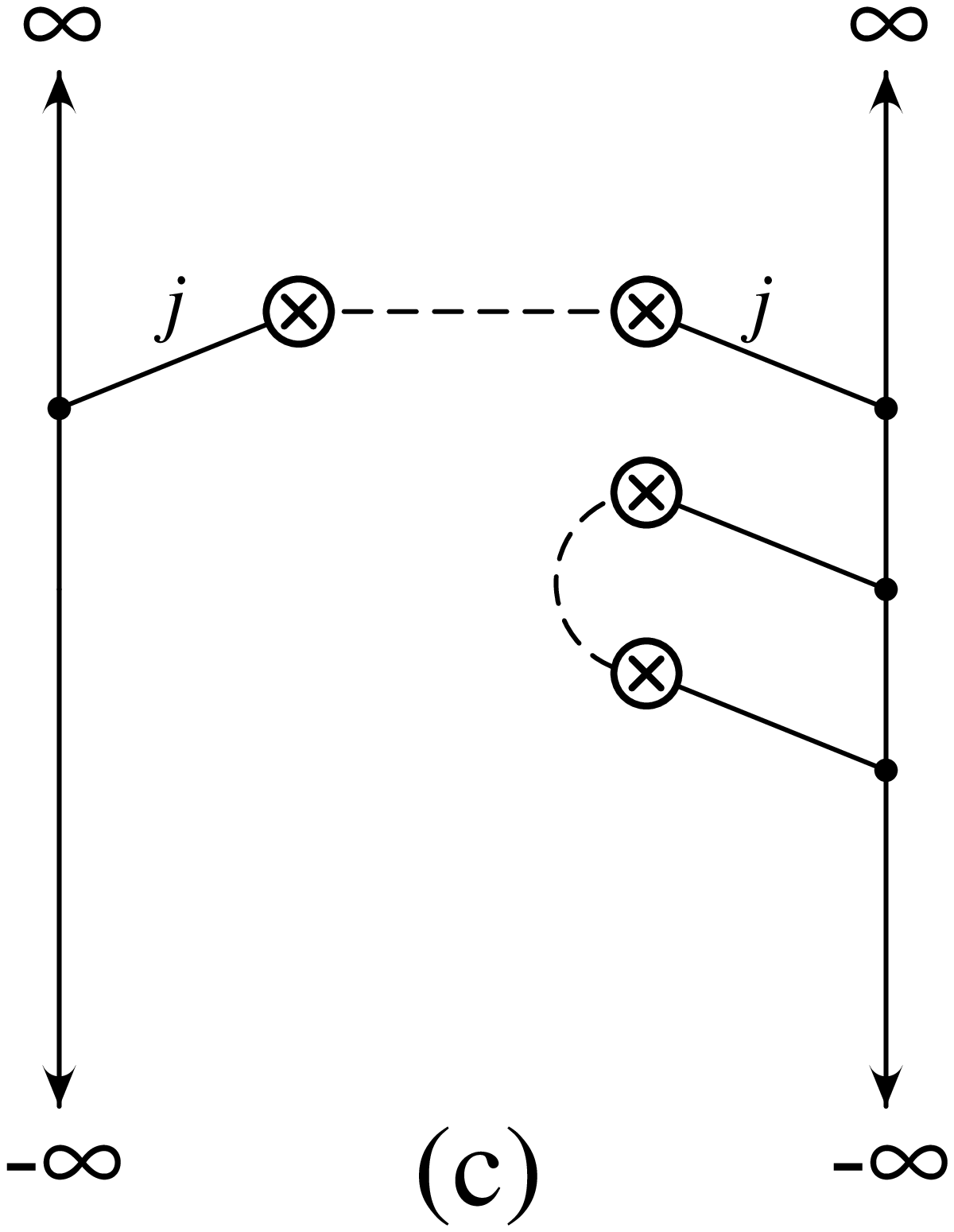}
\includegraphics{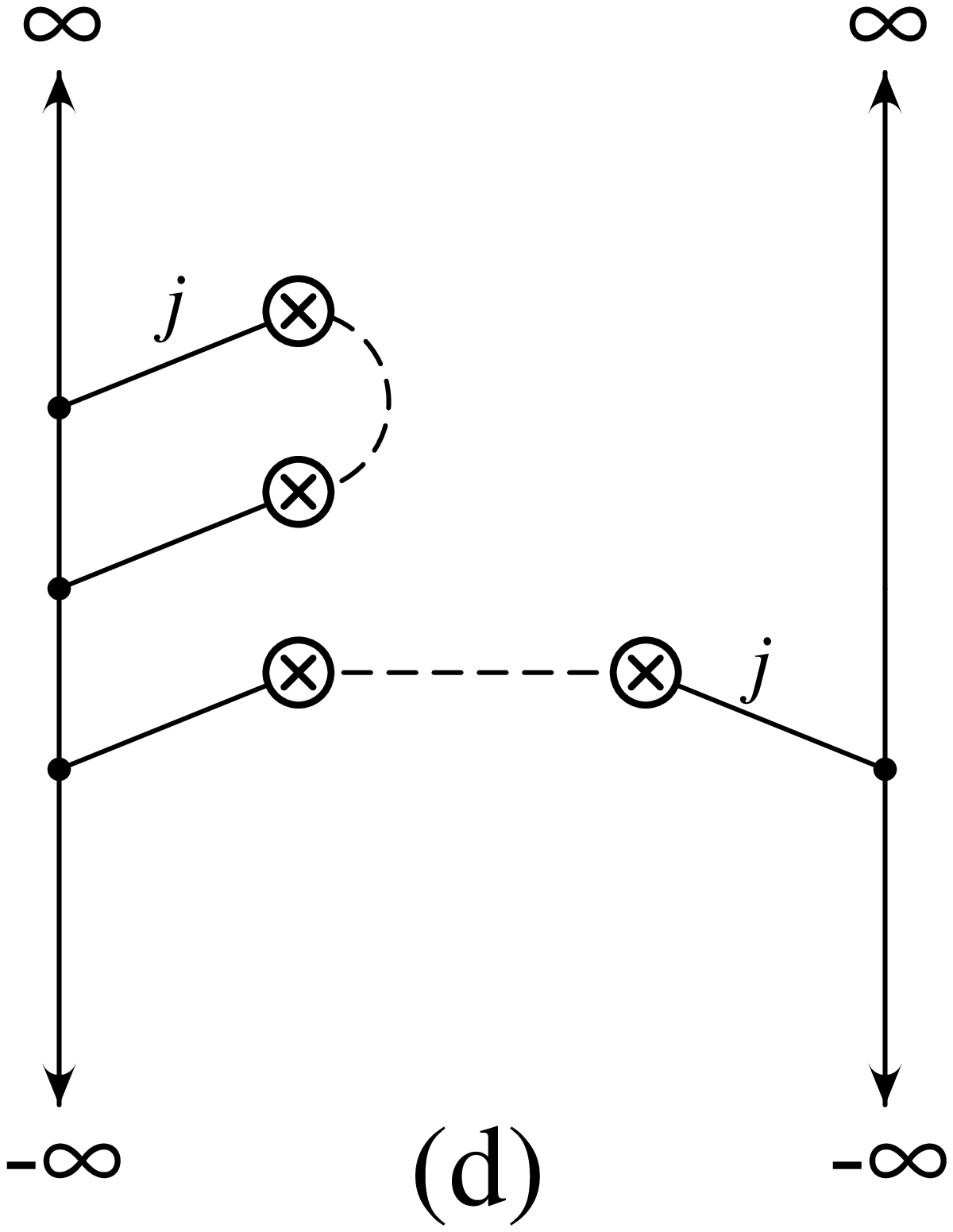}
\includegraphics{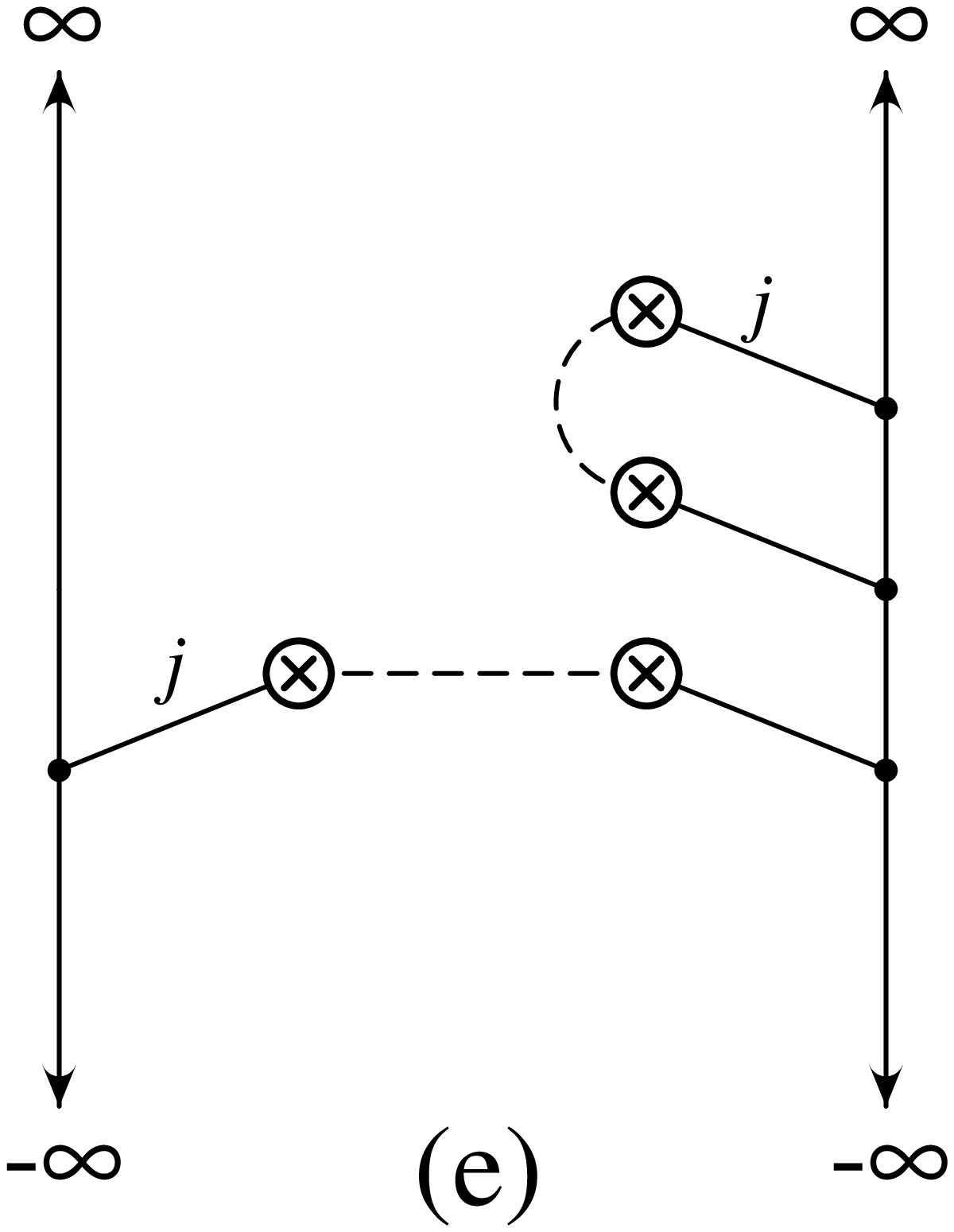}
\includegraphics{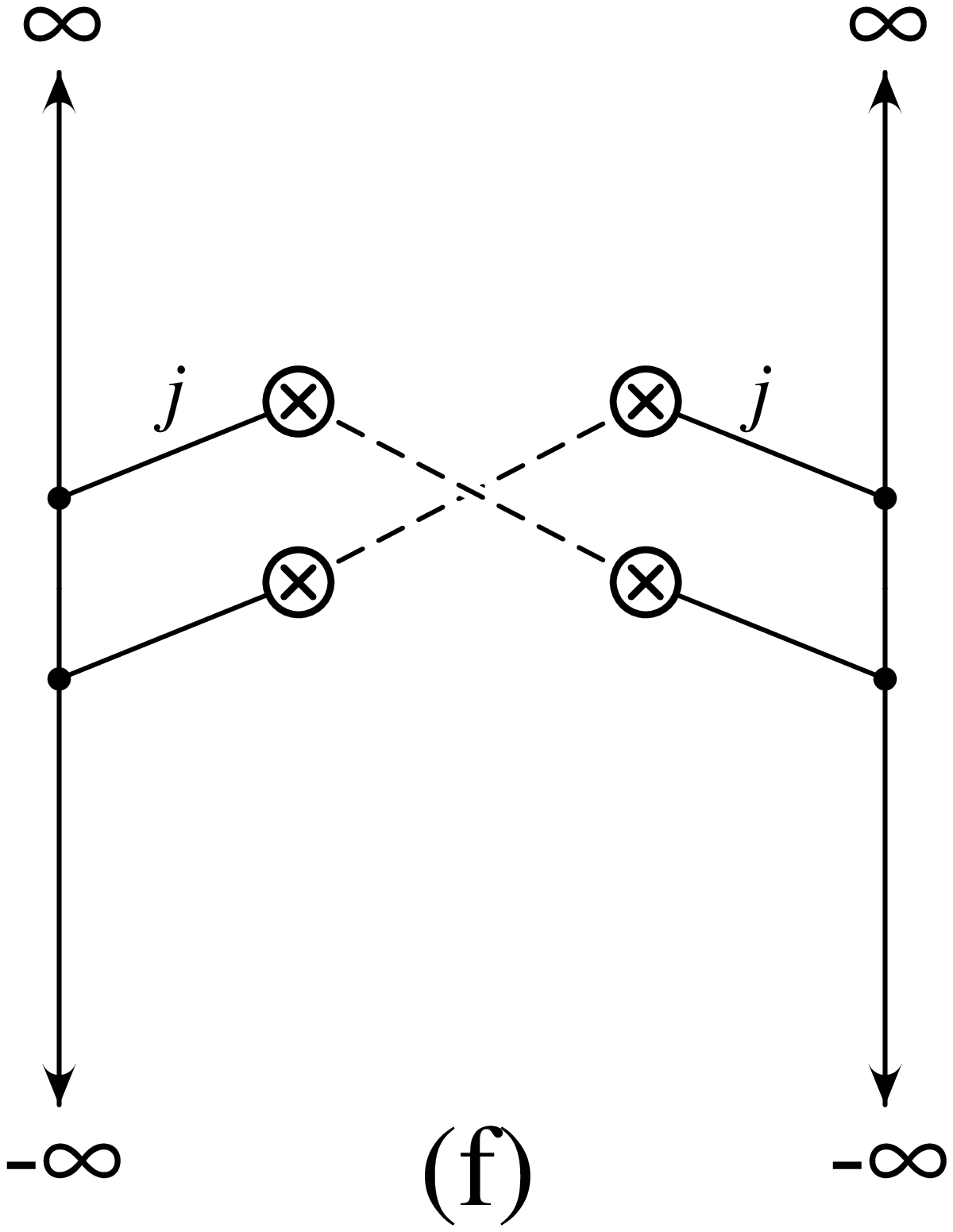}
\includegraphics{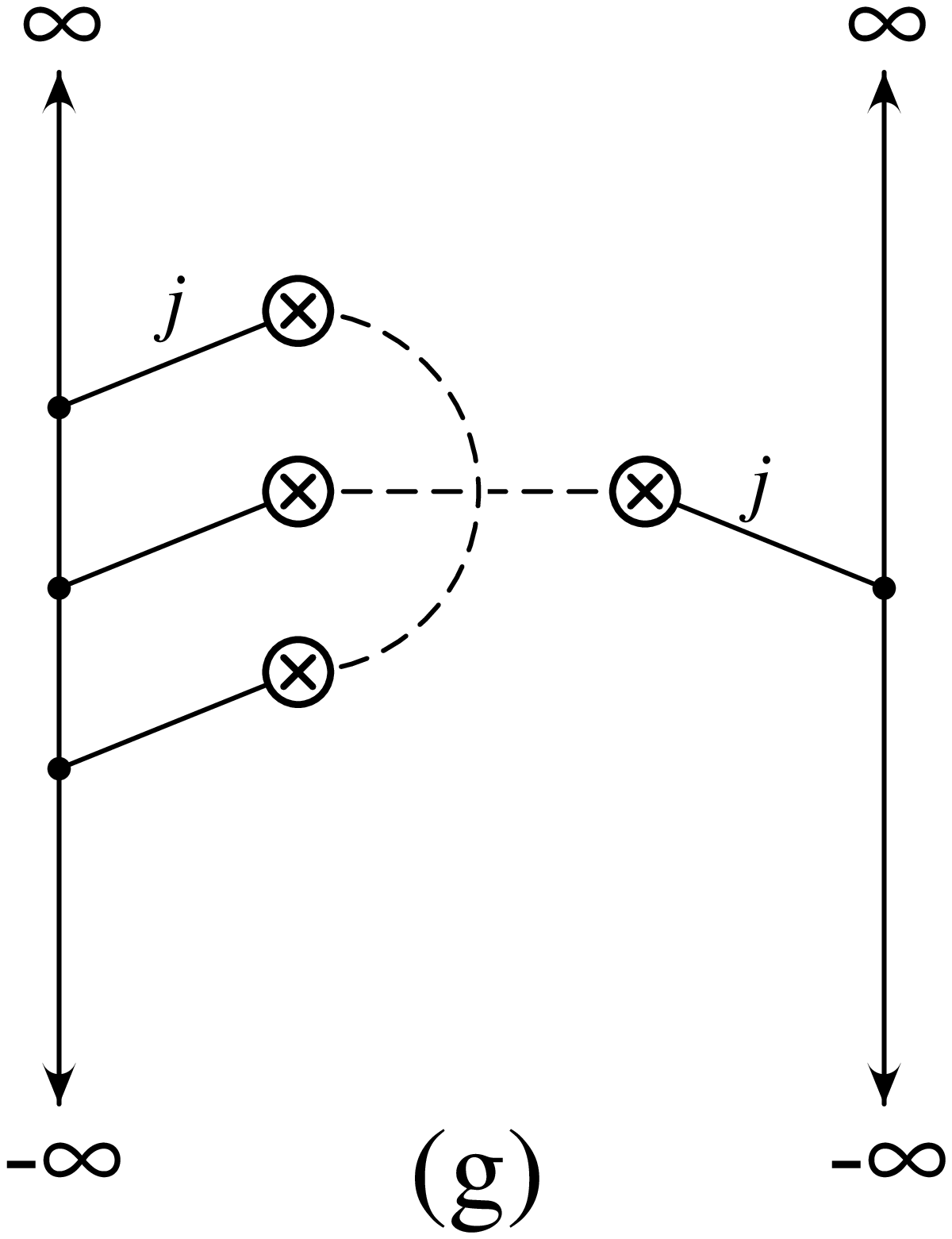}
\includegraphics{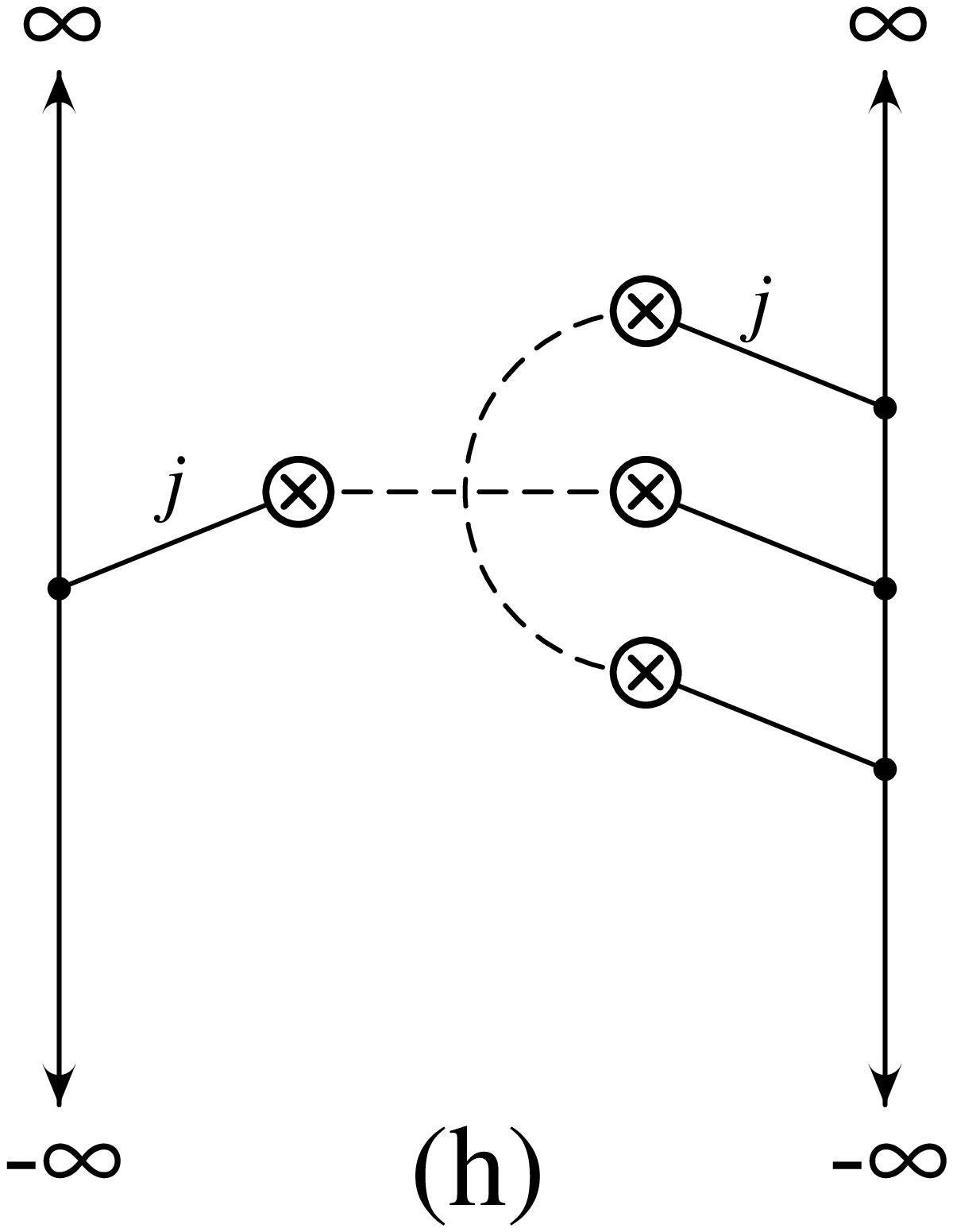}
\includegraphics{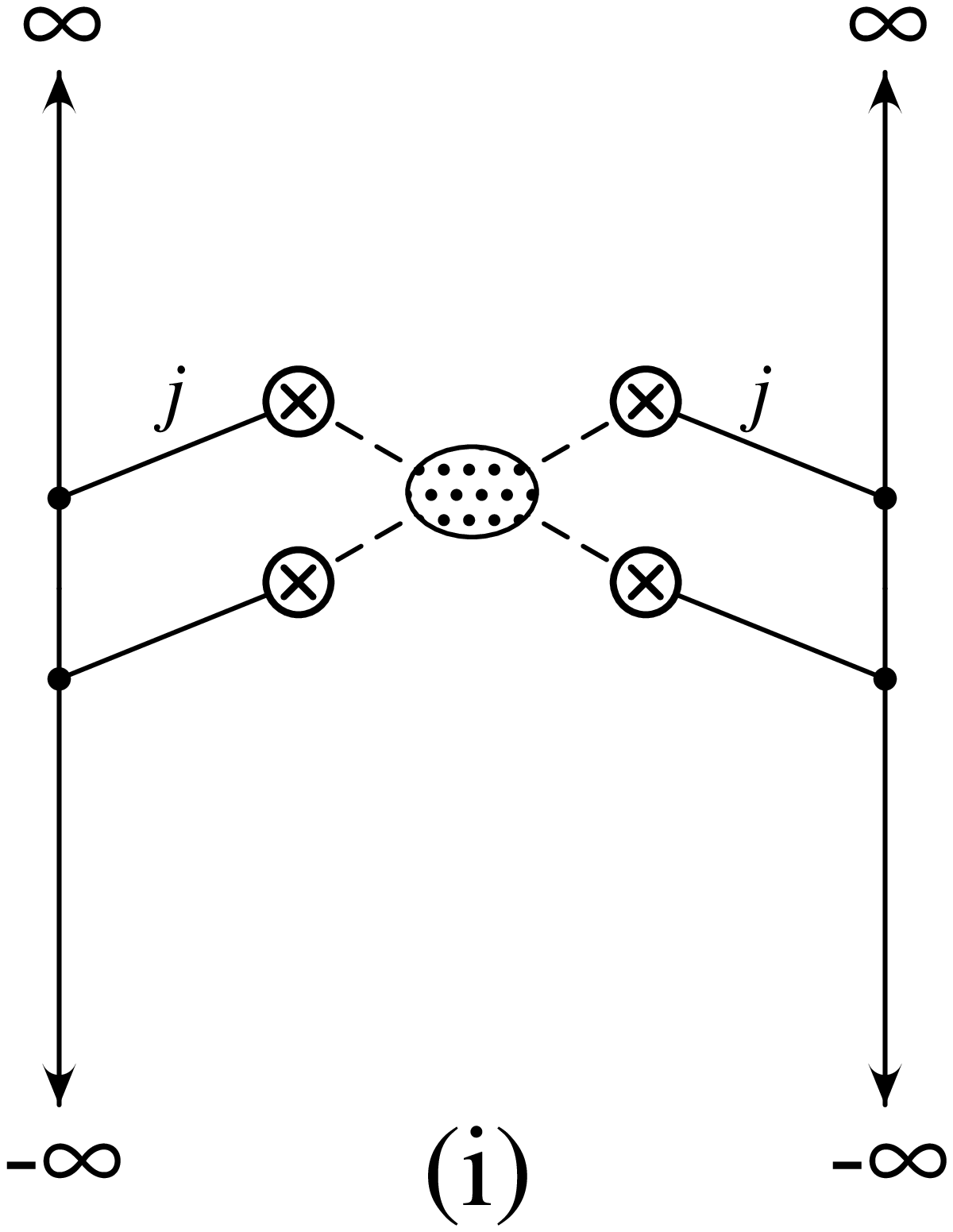}
\includegraphics{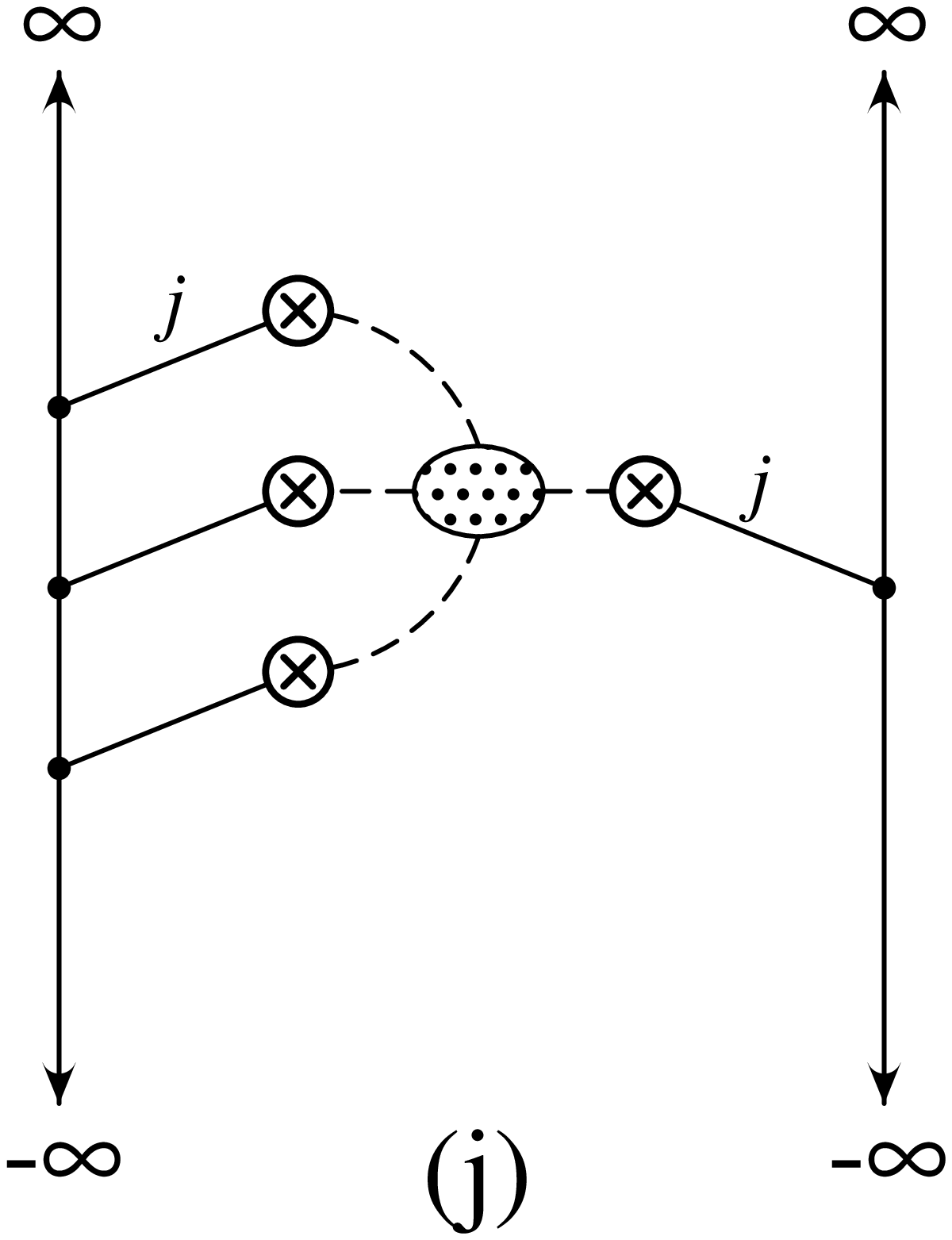}
\includegraphics{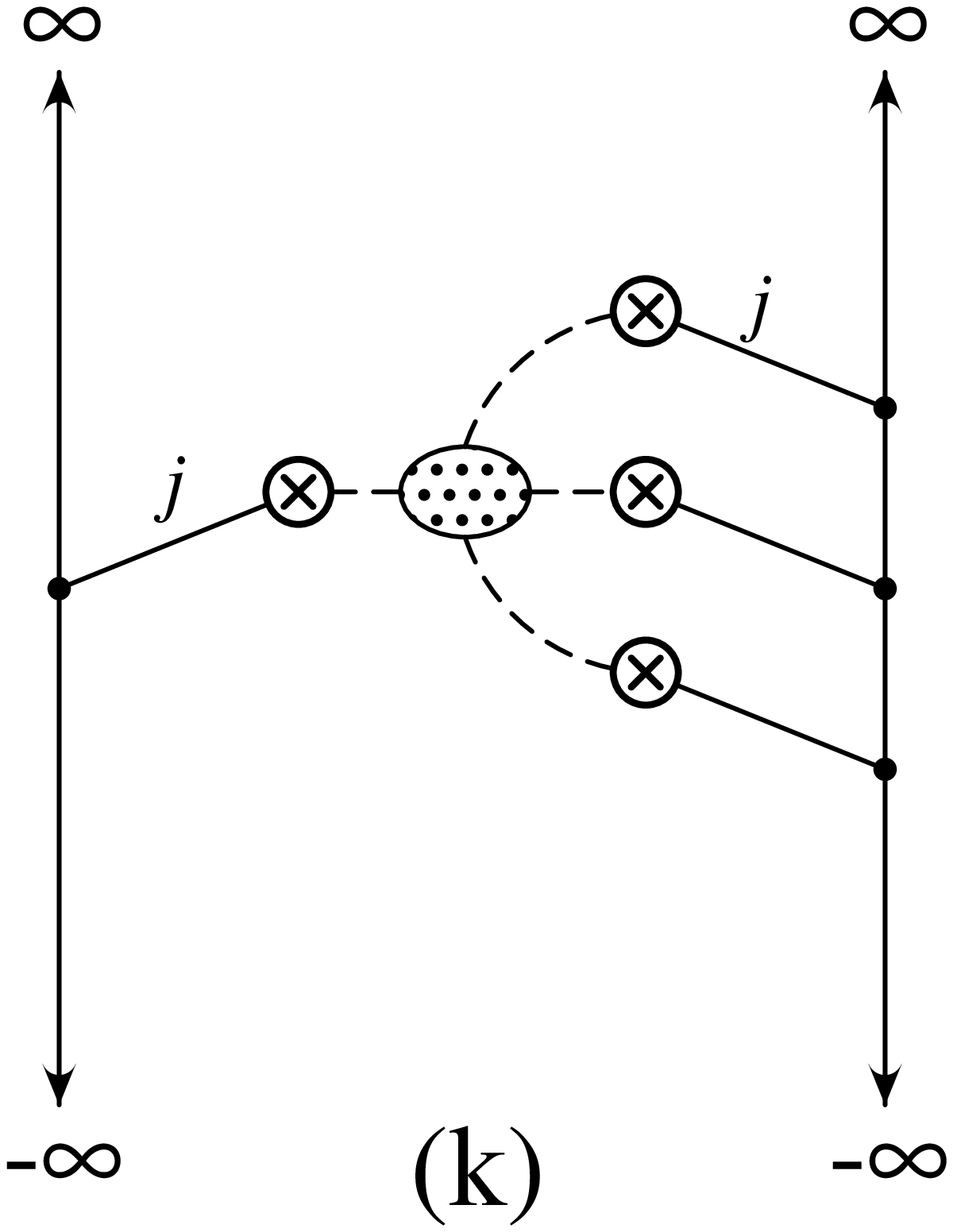}

\caption[]{The contributions to the gluon number 
density~(\protect\ref{gnum3d})
at 4th order in $\rho$.  Diagrams (f)--(h) are non-planar,
whereas
diagrams (i)--(k) represent
the (non-Gaussian) contribution from an irreducible 4-point
charge density correlation function.
According to the power-counting rules
described in the text, diagrams (a)--(e) are proportional to 
$\alpha_s^3 A^{4/3}$ whereas diagrams (f)--(h) are proportional
to $\alpha_s^3 A$.
}
\label{FourthOrder}
\end{figure}

To first approximation, the function ${\cal S}_m$ in 
Eq.~(\ref{multi-rho}) is essentially a
constant when we stay well inside
the large nucleus since the individual nucleons are 
identical insofar as the strong interaction is concerned.
Thus, the quantity
$\langle A_i^a(\qz;\xt) A_i^a(-\qz;\xprimet) \rangle$
computed from the expansion given in Eq.~(\ref{LCA})
is of the form
\beq
(R/a)^{c} \ts F(\qz;\xt-\xt')
\sim (A^{1/3})^{c}\ts F(\qz;\xt-\xt')
\label{special-form}
\eeq
where $F$ is a dimensionless function.
For a given diagram, the power $c$ is equal to the number
of independent clusters formed when the vertical lines are
removed.  
Two clusters are independent
if they can be represented by a planar diagram (no crossings).
For example, $c=1$ for
the diagrams in Figs.~\ref{SecondOrder}, \ref{ThirdOrder},
and~\ref{FourthOrder}(f)--\ref{FourthOrder}(k), whereas
$c=2$ for the diagrams in 
Figs.~\ref{FourthOrder}(a)--\ref{FourthOrder}(e).

The origin of the factor of $A^{1/3}$ for each independent
cluster is the divergence which appears in some of the integrations
when we take $\SIII_m = const.$:  these are the integrations which
get cut off at the scale $R$ instead of the scale $a$.
To see this, imagine that we have changed the integration
variables to a set of sum and difference coordinates.
The sum (center-of-mass) coordinates
describe the positions of the clusters (whether or not 
these clusters are independent).\footnote{Note that transverse
position $(\xt+\xt')/2$ of the clusters is not integrated over when
computing $\langle A_i^a(\qz;\xt) A_i^a(-\qz;\xprimet) \rangle$.
Hence, the only center-of-mass integrations present are
longitudinal in nature.}
The difference coordinates describe cluster-cluster separations
as well as the internal separations of the components of each cluster.
All of the difference integrals are finite
and so inherit the scale $a$ associated with $\DIII_m$.
On the other hand, the center-of-mass integrations are insensitive
to $a$:  so long as the cluster is located somewhere inside the
nucleus, the integrand can be significant.  Hence, these integrations
produce factors of $R$ instead of $a$.  
In the absence of the ordering of the locations of the points on
the vertical lines, then, we would obtain a factor $(R/a)^{k}$
for $k$ clusters.  What the ordering does is to force clusters
which cross each other in nonplanar diagrams to maintain a relative
separation which does not greatly exceed $a$.  Thus, if only $c$
clusters are free to move independently of each other, the factor
we obtain will be only $(R/a)^c \sim (A^{1/3})^c$.

Finally, we note that every diagram will pick up an additional
factor $(R/a)^2 \sim A^{2/3}$ when the Fourier transform indicated
in Eq.~(\ref{gnum3d}) is performed.  The origin is the same as
above:  the integration over $\xt{-}\xprimet$ gets cut off at
the scale $a$ whereas the integration over $(\xt{+}\xprimet)/2$ is
cut off at the scale $R$.  Assembling all of the pieces, we have
the result that the diagram containing $j$ sources and $c$ independent
clusters contains the factor $\alpha_s^{j-1} (A^{1/3})^{c+2}$.

With our power-counting rules in hand, we may now classify
the relative contributions of the diagrams in 
Figs.~\ref{SecondOrder}--\ref{FourthOrder}.
The lowest order diagram is given in 
Fig.~\ref{SecondOrder}, and contains two sources and a single
cluster.  Hence, its contribution is of order $\alpha_s A$.
The third-order diagrams of Fig.~\ref{ThirdOrder}
have a total of three sources tied together as a single cluster.
Their contribution is therefore order $\alpha_s^2 A$ and
is suppressed by a power of $\alpha_s$ relative to the leading
order diagram.
Finally, let us look at the fourth-order diagrams 
of Fig.~\ref{FourthOrder}. According to our power-counting
rules, diagrams (a)--(e) all contribute at the $\alpha_s^3 A^{4/3}$
level (four sources and two independent clusters),
whereas diagrams (f)--(k) produce only $\alpha_s^3 A$.
We imagine that we have a very large nucleus such that
$\alpha_s^2 A^{1/3}$ is of order 1 or greater.
Thus, we consider diagrams (a)--(e)
to be as important as the leading order, whereas diagrams (f)--(k)
are subleading, being suppressed by one factor of $a/R \sim A^{-1/3}$.

The generalization to higher orders is obvious.  Clearly, the 
leading diagrams are all ladders:  they contain only two-point
contractions which do not cross, thus producing the maximum number
of independent
clusters for a given number of sources.
All other diagrams are suppressed:
either they contain 
extra factors of $\alpha_s$ or 
they have fewer factors of $A^{1/3}$.
Therefore, the $A$ and $\alpha_s$ dependence of the leading
contributions to the
final result must look like
\beq
\xf\ts\gnumX = c_0
\alpha_s A
\Bigl( 1 + c_1 \alpha_s^2 A^{1/3} 
         + c_2 \alpha_s^4 A^{2/3} 
         + \cdots \ts\Bigr).
\label{expand}
\eeq
But this is exactly the form obtained by expanding the 
expressions contained in Eqs.~(5.18)--(5.20) of Ref.~\cite{PAPER16}.
In fact, the set of diagrams described above is
precisely the set of diagrams considered
in Ref.~\cite{PAPER16}.  Moreover, it is also the set of diagrams
whose two-dimensional reduction corresponds to those terms
which were retained in Refs.~\cite{paper9,PAPER14}.
These contributions are easily resummed into a relatively
simple ``exponential''\cite{paper9,PAPER14,PAPER16}.
Thus, we see that even if the weight function $W[\rho(\xIII\ts)]$
is non-Gaussian, we obtain the same result 
to leading order as if we had chosen to use a Gaussian weight
function instead.


\section{Implications}\label{CONCLUSIONS}

We have just demonstrated that the non-Gaussian portions
of the statistical weight function $W[\rho(\xIII\ts)]$ in
the MV model do not contribute to the gluon number density
at leading order.  On the other hand, the results of Ref.~\cite{paper91}
suggest that the quantum corrections to the MV model
may be incorporated by solving the RGE for the weight function
and using this (non-Gaussian) result to perform the ensemble
averages required to compute the gluon number density.
A synthesis of these two conclusions has the following implication:
it is sufficient to solve the RGE for the new value of the
two point charge-density correlation function, and to use this
renormalized correlator as defining an effective Gaussian weight
function to be input to the MV model.  The separation scale between
hard and soft partons appearing in Ref.~\cite{paper91} may be
recast as a dependence of the renormalized correlator on $\xf$.
In terms of Eq.~(\ref{rhorho3}), this dependence can potentially
show up in the detailed shape of the smooth function 
$C(\xIII-\xprimeIII)$,
as well as in the prefactors which were suppressed in writing
down Eq.~(\ref{rhorho3}).   In terms of the 
expansion~(\ref{expand}), this means that the $c_i$'s would be
$\xf$-dependent.  Since the quantum corrections presumably 
incorporate the physics of the Yukawa cloud believed to surround
the nucleons, it is possible that the detailed relation between
$a$ and $R$ could change somewhat, again in an $\xf$-dependent
way.

In Ref.~\cite{paper9} it was argued that not only should
the RGE-improved MV model apply to nuclei at sufficiently small 
values of $\xf$,
but that it should also apply to hadrons as well, at even smaller
values of $\xf$.
Our argument does {\it not}\
apply to this situation.  For a hadron we would effectively
have $a/R = 1$, meaning that all of the suppressed contributions
we have dropped are no longer unimportant.  In this situation
the full non-Gaussian solution to the RGE has relevance
and must be taken into account.


\acknowledgements

High energy physics research at McGill
University is supported in part by
the Natural Sciences and Engineering Research Council of Canada
and the Fonds pour la Formation de Chercheurs 
et l'Aide \`a la Recherche of Qu\'ebec.
High energy physics research at the University of Arizona
is supported by the U.S. Department of Energy
under contract number DE-FG02-95ER40906.




\begin{table}
\caption{Mapping between the notations used in the discussion
of the central limit theorem and those associated with the
MV model.
\label{Translation}}
\begin{tabular}{cccc}
& central limit theorem  & MV model              & \\ \hline
& $\r_i^{(p)}$           & $\rho^a(\xz;\xt)$     & \\
& $(p)$                  & $\xz$                 & \\
& $i$                    & $\{a;\xt\}$           & \\
& $N$                    &  $A^{1/3}$            & \\
& $\m_i \equiv {1\over N} \sum_{p=1}^{N} \r_i^{(p)}$ 
& $A^{-1/3}\rho_2^a(\xt) \equiv 
   A^{-1/3} \int_{-\infty}^{\infty} d\xz\ts\rho^a(\xz,\xt)$    & \\[0.1cm]
& $W(\v\m)$              &  $W[\rho_2(\xt)]$     &
\end{tabular}
\end{table}

\end{document}